\documentclass[conference]{IEEEtran}
\IEEEoverridecommandlockouts

\usepackage[utf8]{inputenc}
\usepackage{amsmath}

\usepackage{mathtools}
\usepackage{amssymb}
\usepackage{listings}
\usepackage{array}
\usepackage{xspace}
\usepackage[normalem]{ulem} 
\usepackage{graphicx}
\graphicspath{ {./figures/} }
\usepackage{varwidth}
\usepackage[ruled, linesnumbered]{algorithm2e}
\usepackage{enumitem}
\usepackage{footnote}
\usepackage{tablefootnote}
\usepackage{tabularx}
\usepackage{multirow}
\usepackage{xcolor}

\usepackage{url}

\usepackage{caption}

\newif\ifsubmission
\submissiontrue 

\ifsubmission

\newcommand{\rev}[1]{#1}

\fi

\newcommand{\sys}{DSLog\xspace}

\begin{document}

\title{Compression and In-Situ Query Processing for Fine-Grained Array Lineage}

\author{JinJin Zhao, Sanjay Krishnan\\
\IEEEauthorblockN{ChiData, University of Chicago}
\texttt{\{j2zhao,skr\}@uchicago.edu}\\
}

\SetKwFunction{FRange}{RangeGroup}
\SetKwFunction{FOut}{OutputEqual}
\SetKwFunction{FIn}{InputEqual}
\SetKwFunction{FRel}{AddRelative}
\SetKwFunction{FMainS}{ProvRC-S}
\SetKwFunction{FMain}{ProvRC}
\SetKwProg{Fn}{Function}{:}{}

\maketitle
\begin{abstract}
Tracking data lineage is important for data integrity, reproducibility, and debugging data science workflows.
However, fine-grained lineage (i.e., at a cell level) is challenging to store, even for the smallest datasets.
This paper introduces \sys, a storage system that efficiently stores, indexes, and queries array data lineage, agnostic to capture methodology. 
A main contribution is our new compression algorithm, named \FMain, that compresses captured lineage relationships.
Using \FMain for lineage compression result in a significant storage reduction over functions with simple spatial regularity, beating alternative columnar-store baselines by up to \rev{2000x}. We also show that \FMain facilitates in-situ query processing that allows forward and backward lineage queries without decompression - in the optimal case, surpassing baselines by \rev{20x} in query latency on random \texttt{numpy} pipelines. 
\end{abstract}

\section{Introduction} \label{sec:introduction}
As data scientists work with ever more complex data workflows, data lineage is a vital tool to ensure the reproducibility and reliability of results. 
Data lineage systems represent input-to-output data relationships caused by applying some data manipulation operation on one or more datasets.
After a series of operations, forward queries in a lineage system identify all final output data influenced by some initial input, and backward queries identify all initial input data relevant to some final output.
However, answering such queries in a purely posthoc way can be as expensive as running the original program --- consider running a lineage query on an overnight reporting job in a data warehouse.
Thus, most practical lineage systems construct some form of a \emph{lineage materialization} to improve performance~\cite{psallidas2018smoke}.
This materialization captures key input-to-output relationships to speed up future forward and backward queries.

For fine-grained lineage, these materializations quickly hit a problem of scale. 
An uncompressed materialization for a point-wise summation of two matrices of 1M cells each can be 10s of MBs (see Section \ref{sec:experiments}).
In other words, the stored lineage may often be larger than the data itself.
Fortunately, this data is often highly structured and repetitive.
The critical insight of this paper is that particular lineage structures can be exploited to build storage-efficient compressed lineage tables and fast query strategies over these tables that do not require decompression.
While it might be obvious that lineage materializations can be effectively compressed, building compressed lineage tables that can be queried in situ is an open technical challenge addressed in this paper.
We contribute \sys: a system that facilitates the efficient storage, query, and reuse of fine-grained array-like lineage.
\sys is an indexing service that stores lineage relationships captured by lineage capture frameworks. Its API ingests lineage for data science workflows in which the data is structured in multidimensional arrays (or data types resembling arrays). Integer-indexed multidimensional arrays are a universal interface in data science that can represent data frames, documents, time series, and images. Over the relationship captured in \sys, users can efficiently make two types of queries across multiple workflow steps: forward queries that find all output data contributed to by an input and backward queries that find all input data contributing to an output. 

Several architectural insights are contributed here: (1) individual lineage relationships are automatically encoded in a novel compressed representation; (2) the query processing of forward and backward queries is in situ over the compressed representation, and (3) the captured lineage is stored in a normalized relational format with ``index reshaping'' that can be reused for future calls to the same operation. The basis of these architectural insights is the \FMain lineage compression algorithm. This algorithm is used in \sys to exploit the high degree of spatial coherence between input and output indices of common data science operations, using a novel combination of multi-attribute range encoding and relative indexing techniques.

In the implementation,  we integrate \sys with the \texttt{Python} data science stack. We demonstrate experiments where the prototype captures fine-grained lineage capture with arrays of up to a million cells. The \FMain algorithm achieves a considerable storage reduction over operations with simple spatial regularity. It achieves an approximately 0.3\% storage ratio over raw storage and beats alternative columnar-store baselines by up to 1400x, with minimal overhead. We also demonstrate significant performance improvements in the downstream query processing, performing queries over 100,000 input cells in $<1$ second in the optimal case, beating the same baselines by up to \rev{1500x}. Finally, we show that the lineage reuse capability enabled by this compressed representation has high coverage over the \texttt{numpy} library, achieving input-independent reuse on 99 of the 136 evaluated \texttt{numpy} operations.

\section{Background} \label{sec:background}
First, we motivate our problem setting and describe the related work.

\subsection{Problem Setting and Scope}
An \emph{array} of data is a rectangular data collection indexed by tuples of positive integers. 
Each value in an array is called a \emph{cell}. The cells can be fully indexed by their position along each array \emph{axis}.
The length of the tuples indexing an array defines its \emph{shape}.
Arrays can represent any ordered data structure and act as a useful universal data type for data science operations.
\begin{itemize}
    \item A dataset of machine learning examples can be represented as a 2D array of feature vectors, with each cell represented by a floating point number.
    \item An image can be represented as a 3D array of RGB pixels, each represented by an unsigned integer 0-255.
    \item A document can be represented as a 1D array of strings.
    \item A relational table can be represented as a 2D array where the rows and attributes are organized in canonical order.
\end{itemize}

An array program is simply an operation that takes as input a set of input arrays and outputs another array. For example,
\begin{lstlisting}
Y = foo(X1,...,XN)
\end{lstlisting}
the operation \texttt{foo} applies to arrays $X^1,...,X^N$ and outputs another array $Y$.

\emph{Fine-grained lineage} for a single operation refers to contribution relationships between input and output cells in such operation: which elements in the input array contribute to the values of each output element. For any individual operation, we assume that we have a capture method where, given a cell in an output array of an operation, we can find all the cells from the input arrays that ``contributed'' to that output cell. Many pre-existing prototype systems capture and generate some form of lineage for data science \cite{suh2004secure, tiwari2009complete, vartak2018mistique, xin2018helix, sellam2019deepbase, rezig2020dagger, shang2019democratizing,zhang2017diagnosing, phani2021lima, nikolic2014linview, boehm2016systemml}. \sys allows for the incorporation of this lineage regardless of capture methodology. In Section \ref{sec:experiments}, we present our implementation of different types of these systems to generate such lineage for evaluation of \sys. \sys work perpendicular to individual operation lineage capture to efficiently store, query, and reuse lineage across multiple operations.

\subsection{Architecture}

Next, we describe the high-level architecture of \sys and where it sits on the data science lineage workflow. When users create array-based data science pipelines, fine-grained lineage is captured based on the specified capture techniques. While \sys does not provide a complete solution for capture, it does support capture in cases where the lineage can be reused. Captured lineage is stored in a compressed format within \sys using \FMain, and users can issue forward and backward queries over this to get information about their workflow across multiple operations. \sys responds to the queries with an in-situ algorithm over the stored compressed lineage. In this way, storage and queries are fully supported.

 \section{\sys: An Efficient Lineage Management System} \label{sec:system}
\begin{figure}[t]
\tiny
\centering
\begin{tabular}{c}
\textbf{(A) Python Code Representation} \\
\begin{lstlisting}
A = numpy.array([[0,3],[1,5],[2,1]])
B = numpy.sum(A,axis=1)
\end{lstlisting}
\end{tabular}
\begin{tabular}{ccc}
\multicolumn{3}{c}{\textbf{(B) Relational Representation}}\\\hline
$b_1$ & $a_1$ & $a_2$ \\\hline
\textbf{1} & \textbf{1} & \textbf{1} \\
1 & 1 & 2 \\
2 & 2 & 1 \\
2 & 2 & 2 \\
3 & 3 & 1 \\
3 & 3 & 2 \\ \\
\end{tabular}

\begin{tabular}{c}
{\textbf{(C) Lineage of First Row of Relational Representation}}\\
\includegraphics[width=.3 \linewidth]{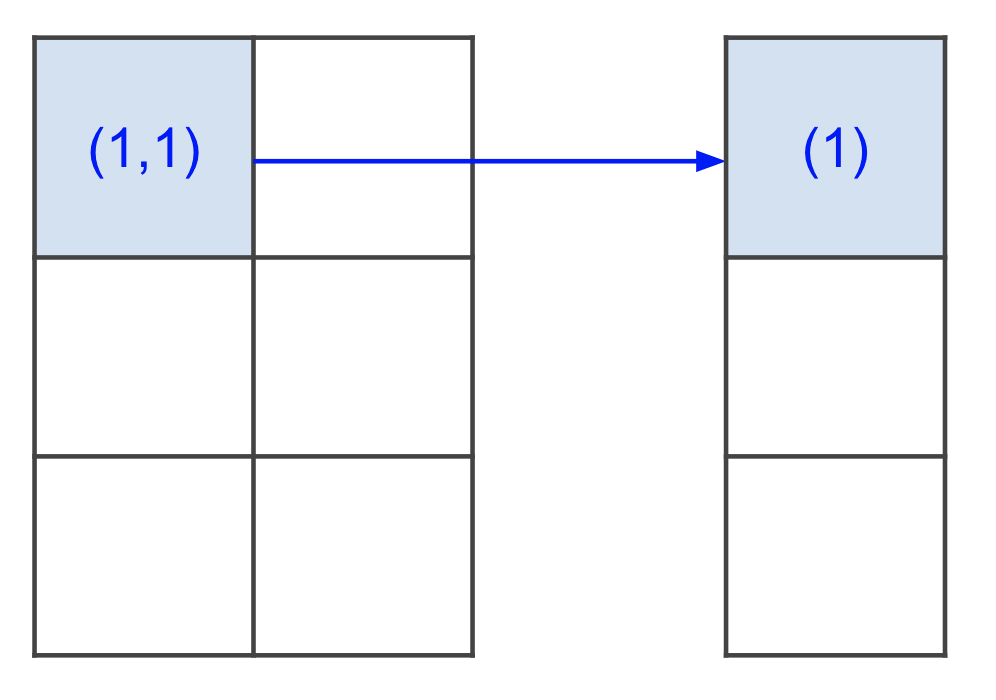}
\end{tabular}

\vspace{5pt}
\caption{We present a typical operation that sums over the second axis of an array (A), the relational representation of such operation (B), and an array visualization of the lineage relation described by the first row of the relational representation (C).}
\label{fig:cell2cell}
\end{figure}
\sys is a system that indexes input-to-output lineage relationships. We describe the API and the basic data model in this section.

\subsection{\sys API}
First, we present the basic API that allows an application developer to connect a lineage system with \sys.

\lstset{%
    basicstyle=\ttfamily\footnotesize,
    breakatwhitespace=false,
    breaklines=true,
    captionpos=b,
    commentstyle=\textcolor{red},
    escapeinside={\%*}{*)},
    extendedchars=true,
    frame=none,
    keepspaces=true,
    keywordstyle=\textcolor{black},
    language=Java,
    numbers=none,
    showspaces=false,
    showstringspaces=false,
    showtabs=false,
    aboveskip=5pt,
    belowskip=5pt,
    sensitive=false
}

\vspace{0.5em} \noindent \textbf{Defining Tracked Arrays. } \sys allows users to define named Arrays with a specified shape. These are the arrays that are tracked through a data science workflow.
\begin{lstlisting}
Array(name: String, shape: Integer[])
//defines an array with the specified shape
\end{lstlisting}
\vspace{0.5em} \noindent \textbf{Defining Lineage. }  We can capture the lineage of these Arrays with a lineage object. 

\begin{lstlisting}
Lineage(arr1: String, arr2: String, capture: Object)
//defines the lineage relationship between two arrays
//capture(i: Long[]) returns cells in arr2 that are linked to the cell in arr1 at index i
\end{lstlisting}

The user specifies a capture method to record lineage between \texttt{arr1} and \texttt{arr2}. A capture method generates an iterator where, for each cell in the output array, we can retrieve the spatial indices (of the cells) in the input array with a contribution relationship. The user's desired capture method will be called internally to populate the lineage object in the storage manager.

\sys enables efficient lineage capture for multiple arrays operated over one operation. 

\begin{lstlisting}
register_operation(op_name: String, in_arrs: String[], out_arrs: String[], capture: Object[], op_args: Any[], reuse: Optional Boolean)
//registers an operation where provenance is captured between input and output arrays
\end{lstlisting}

This is implemented with an operation registration API consolidating lineage within an executed data science operation. This method internally creates lineage between linked input and output array pairs of \texttt{op\_name} and registers a record between those lineage tables and the operation metadata. 

The API enables automatic capture with index reshaping when "reuse" is enabled. In that case, the system tries to generate lineage solely based on previous registrations without requiring a pre-defined capture methodology in future API calls with the same operation name. Since lineage capture is usually expensive, this can significantly reduce the cost of lineage in data science overall. 


\vspace{0.5em} \noindent \textbf{Queries Over lineage. } The user can issue forward and backward queries to the storage manager over Arrays. These queries can be expressed in a single query interface.  

\begin{lstlisting}
prov_query(X: String[], query_cells: Integer[[]])
//retrieves lineage between cells in query_cells of the first Array and the cells in the last Array along the path X
\end{lstlisting}

Based on a query path, $X$, the method \texttt{prov\_query} retrieves the cells in the last Array linked to specified cells, \texttt{query\_cells},  of the first Array. This query path is expressed as a sequence of Arrays, $X: [X^1, \dots X^n]$. For example, consider the operations \texttt{Y = foo(X)} and \texttt{Z = bar(Y)}. One possible query path is $[X, Y, Z]$, representing a forward path from the cells of $X$ to that of $Z$. Another possible query path is $[Z, Y, X]$, representing a backward path from the cells in $Z$ to those of $X$. If we query along these paths, we can return fine-grained lineage between the cells in $Z$ and the cells in $X$.

\subsection{Relational Data Model}\label{sec:rel}
\sys captures lineage in a relational data model.
Every contribution lineage relationship is a mapping between input and output array cells \cite{chapman2008efficient,olteanu2011factorisation,interlandi2015titian, widom2004trio}. Such a mapping is elegantly represented as a relation. 
Let $A$ be an m-dimensional input array whose axes are denoted with $a_1, a_2,...,a_m$. Similarly, the output array, $B$, is an l-dimensional array with axes denoted with $b_1, b_2,...,b_l$. The lineage relationships between $A$ and $B$ can be represented in a relation over the dimensions $R(b_1, b_2,...,b_l, a_1, a_2,...,a_m)$. Each row in the relation corresponds to a single output lineage $B[b_1, b_2,...,b_l] \leftarrow A[a_1, a_2,...,a_m]$. 

As a concrete example, let us consider a simple array operation over an array with shape 3x2: \texttt{B $=$ numpy.sum(A, axis = 1)}, as shown in Figure \ref{fig:cell2cell}. This operation uses the \texttt{numpy} syntax and sums over one of the axes of the array (A). 
A table representation of the lineage of this operation can be made with \sys (B). Each attribute of the table represents an axis of the input or array, and each value of the table represents an index on that axis. Therefore, each row defines two cells using the axis indices - one in the input array and one in the output array. The rows represent lineage relationships between input and output cells. For example, the first row shows that the element at index \rev{(1,1)} in the input contributes to the element at index $1$ in the output (C).

This format allows for compositional forward and backward queries that span multiple operations.
The lineage relationship over multiple operations can be reconstructed from these tables with a natural join operation ~\cite{interlandi2015titian}. For example, given $R^1$ is the lineage of the operation $A \xrightarrow{} B$ and $R^2$ is the lineage of $B \xrightarrow{} C$, we can find the lineage of $A \xrightarrow[]{} C$ from $R^1 \bowtie R^2$.
 With relational tables, we achieve data independence; a user can still request forward and backward lineage queries across multiple tables independent of exactly how the data are organized within those tables. 
\section{Compression Algorithm}
 \label{sec:compression}
\sys stores a collection of ingested input-and-output lineage relationships in a compressed format \rev{that exploits contiguous runs of integer indices}. We will be using the lineage table from Table \ref{fig:cell2cell} (B) as the running example for each step of this algorithm. 

\rev{\noindent \textbf{Notation. } 
\begin{itemize}
    \item $R$ denotes a table storing the lineage (as described in Section III.B) for a relationship between arrays $A \to B$. 
    \item Lower case $r$ will be used to denote a row from $R$.
    \item $R$ has attributes \[\Omega_R = \Omega_A \cup \Omega_B = \{b_1, b_2,...,b_l, a_1, a_2,...,a_m\}.\]
    \item The $.$ notation is used to reference a particular attribute. 
\end{itemize}}

\subsection{Lineage Compression Algorithm}
The \FMain algorithm has two main subroutines:
\begin{itemize}
    \item Multi-Attribute Range Encoding. Express the relation as a union of multidimensional ``ranges''.
    \item Relative Value Transformation. Expresses an input attribute numerically relative to an output attribute.
\end{itemize}

\subsubsection*{Step 1. Multi-Attribute Range Encoding over Inputs}
The first step compresses \rev{rectangular ranges} over the attributes corresponding to the input array cells. This step automatically compresses lineage that match the pattern \textbf{(1)}. We use a generalized range encoding form over the input axis columns. Range encoding has been used in previous compression techniques ~\cite{pibiri2020techniques}. The basic idea is, given a set of integers, one can represent the set as a union of ranges, e.g.,
\[\textsf{range}(\{1,2,3,4,9,12,13,14,15\}) = \{[1,4],[9],[12,15]\}.\]

\begin{figure}
\tiny
  \centering
  \begin{tabular}{c}
{\textbf{(A) Lineage of an Aggregate Operation}}\\ \\
\includegraphics[width=.45 \linewidth]{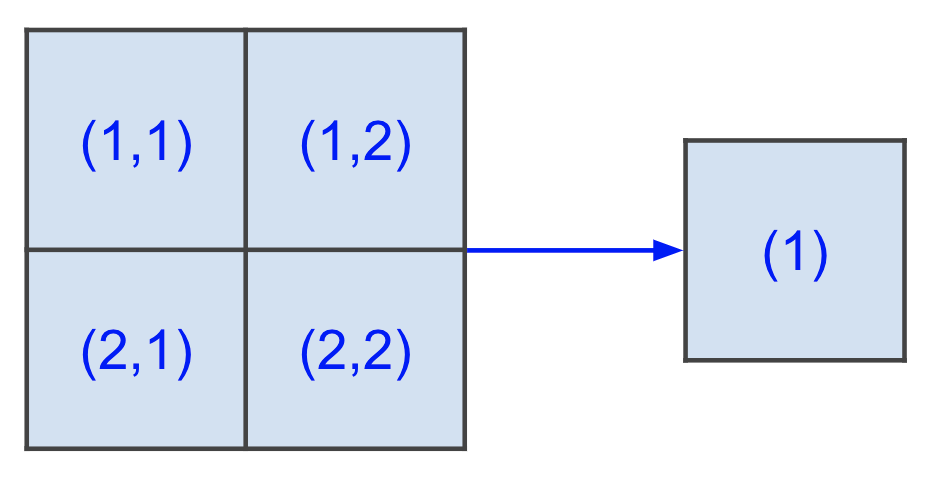}
\end{tabular}
\begin{tabular}{c}
{\textbf{(B) Multi-Attribute Range Encoding Reduction}}\\ \\
\includegraphics[width=.3 \linewidth]{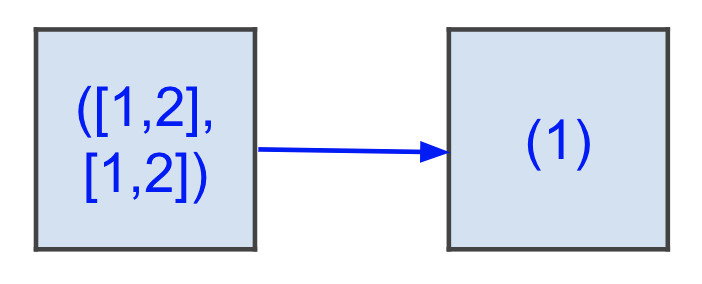}
\end{tabular}
    \caption{This figure shows an example of multi-attribute range encoding, where the all-to-all relationship between four cells in one array and one cell (A) in another array can be represented succinctly with ranges (B).}
    \label{example:range}
\end{figure}

We extend this basic principle to the multi-attribute setting. We iteratively encode each attribute in $R$ into ranges. This algorithm compresses \rev{rectangular ranges} over attributes of array $A$. Figure \ref{example:range} demonstrates this method over a simple array operation that aggregates a 4x4 array into a 1x1 array (A). With multi-attribute range encoding, the input array's indices can be represented as ranges, and we can encode the relationship between those ranges and the output array (B).

First, $R$ is sorted in numeric order over attributes in order of $b_1, b_2,...,b_l, a_1, a_2,...,a_m$. We then first apply the multi-attribute range encoding for input attributes. For each input attribute $a_i \in \{a_m, \dots a_1\}$ and $r \in R$, we add the row into a range, $S$, and then merge the set:

\begin{itemize}
    \item If $S = \emptyset$, set $S: \{r\}$.
    \item Else if $S \cup \{r\}$ is contiguous on $a_i$, set $S : S \cup \{ r\}$.
    \item Else, replace all rows in $R$ that are contained in $S$ with one row $r'$, s. t. $r'.\omega = r.\omega$ for $\omega \in \Omega_R \setminus \{a_i\}$ and $r'.a_i = [low, high]$. Then, set $S: \emptyset$.
\end{itemize}
We state that a set of rows, $S$, form a \rev{rectangular range}, $[low, high]$ on input attribute $a_i \in \Omega_A$, if, all attributes $\Omega_R \setminus \{a_i\}$ have the same value and the set of $\{s.a_i | s \in S \}$ is equivalent to the range of integers spanning $\{low,...,high\}$ (inclusive on both ends). Table \ref{ex:input-prov} shows the example output of this step over our example lineage table. Notice how this step efficiently reduces the number of rows in the lineage table.

\begin{table}[t]
\footnotesize
\centering
\begin{tabular}{lll}
\hline
$b_1$ & $a_1$ & $a_2$ \\\hline
1 & 1 & 1 \\
1 & 1 & 2 \\
2 & 2 & 1 \\
2 & 2 & 2 \\
3 & 3 & 1 \\
3 & 3 & 2 \\
\end{tabular}
$\rightarrow$
\begin{tabular}{lll}
\hline
$b_1$ & $a_1$ & $a_2$ \\\hline
1 & 1 & [1, 2] \\
2 & 2 & [1, 2] \\
3 & 3 & [1, 2] \\
\end{tabular}
\caption{Example Output of Multi-Attribute Range Compression }
\label{ex:input-prov}
\end{table}

\subsubsection*{Step 2. Relative Value Transformation and Output Range Encoding}
In the second step, we apply a relative value transformation between the input and output attributes and then execute the multi-attribute range encoding on the output attributes. This step automatically compresses lineage that matches both patterns \textbf{(2)} and \textbf{(3)}.

\begin{figure}
\tiny
  \centering
  \begin{tabular}{c}
{\textbf{(A) Lineage of a One-to-One Operation}}\\ \\
\includegraphics[width=.25 \linewidth]{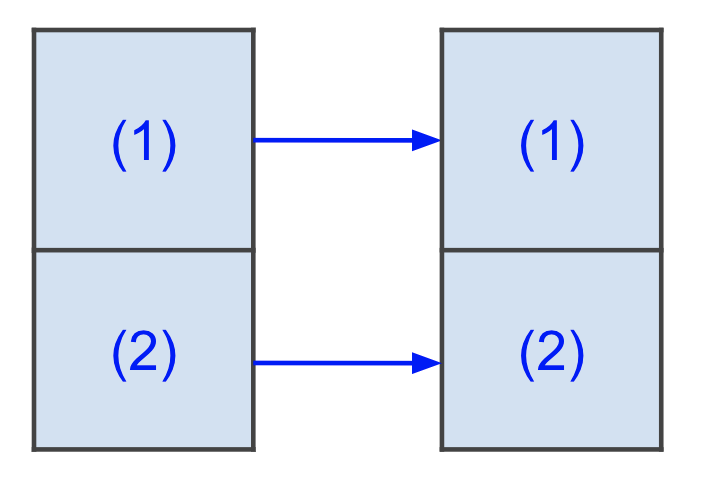}
\end{tabular} 

\begin{tabular}{c}
{\textbf{(B) Relative Transformation}}\\ \\
\includegraphics[width=.25 \linewidth]{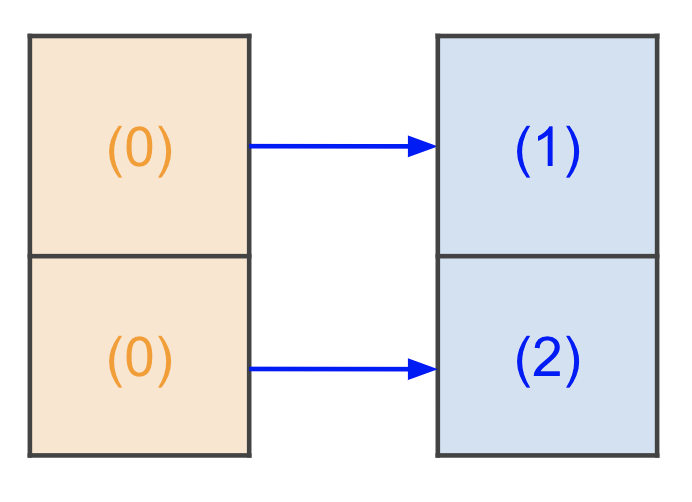}
\end{tabular}
\begin{tabular}{c}
{\textbf{(C) New Range Encoding Opportunities}}\\ \\
\includegraphics[width=.25 \linewidth]{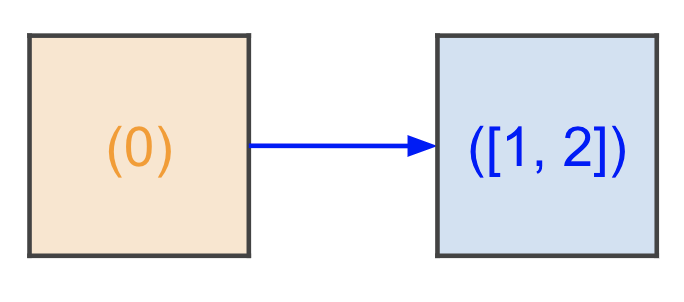}

\end{tabular}
    \caption{This is an example of relative transformation and output range encoding, where, given the one-to-one relationship between two 2x1 arrays (A), we can perform a relative indices transformation on the first array (B), introducing new range compression opportunities for efficient encoding (C).}
    \label{example:relative}
\end{figure}

The relative transformation exploits the cases in lineage relations where, for a set of rows, $S$, and some input and output attributes, $a_i$, $b_j$, for all $s \in S$, we have:

\[s.a_i = s.b_j - \delta \]
\[\delta = s.b_j - s.a_i\]
, for some constant $\delta$. This is observed in many lineage relations - e.g., convolutions, one-to-one operations, matrix multiplication, and rotations. Figure \ref{example:relative} demonstrates this method over a simple one-to-one array operation over a 2x1 array (A). With relative value transformation, the input array's indices are considered relative to the output array's indices (B). This introduces new opportunities for range encoding over the output indices (C).

We perform this algorithm as follows. For every input attribute, $a_i \in \Omega_A$, and output attribute, $b_j \in \Omega_B$, we append a new column to $S$ named $a_ib_j$ that contains the value $b_j - a_i$. If $b_j$ is a range, we subtract $a_i$ from both range values. Our insight is that, for any row $r \in R$ and input attribute, $a_i$, given that we know the output attributes, only one of the values in $a_i, a_ib_1, \dots a_ib_l$ is needed to determine the value of $a_i$.

We apply the same multi-attribute range encoding algorithm over the output attributes $b_j\in [b_l, \dots b_1]$, with simple modifications that account for the above insight. We define that set of rows, $S$, form a range on attribute $b_j \in \Omega_B$ if all attributes $\Omega_B \setminus \{b_j\}$ have the same value, and, for all $a_i \in \Omega_A$, there exists a non-empty subset of $\{a_i, a_ib_1, \dots a_ib_l\}$ where all the attributes have the same value and the set of $\{s.b_j | s \in S \}$ can be represented as a range. When we compress over output attributes by replacing $S$ with a singular row in the multi-attribute range encoding algorithm, we only preserve the values in the non-empty subset rather than every value in $\{a_i, a_ib_1, \dots a_ib_l\}$. If the absolute value is preserved, the lineage matches pattern \textbf{(2)}; otherwise, it matches pattern \textbf{(3)}.

Let us see an example of how this works over the range-encoded example table from the first step (Table \ref{ex:input-prov}). In Table \ref{ex:rel-prov}, the values in $a_1$ are the same as $b_1$ for each row. Adding the column $a_1b_1$ introduces an attribute match in $a_1$ for the multi-attribute range encoding algorithm that did not exist before. We can now compress the previous results to an even more concise representation.

\begin{table}[t]
\footnotesize
\centering
\begin{tabular}{lllll}
\hline
$b_1$ & $a_1$ & $a_2$ &  \\\hline
1 & 1 & [1, 2]  \\
2 & 2 & [1, 2]  \\
3 & 3 & [1, 2]  \\
\end{tabular}
$\rightarrow$
\begin{tabular}{lllll}
\hline
$b_1$ & $a_1$ & $a_1b_1$ & $a_2$ & $a_2b_1$ \\\hline
1 & 1 & 0 & [1, 2] & [0, 1] \\
2 & 2 & 0 & [1, 2] & [-1, 0] \\
3 & 3 & 0 & [1, 2] & [-2, -1] \\
\end{tabular}
$\rightarrow$
\vspace{0.5em}
\begin{tabular}{lllll}
\hline
$b_1$ & $a_1$ & $a_1b_1$ & $a_2$ & $a_2b_1$ \\\hline
[1,3] & - & 0 & [1, 2] & - \\
\end{tabular}
\caption{Output Relative Value Transformation Improving Opportunities for Multi-Attribute Range Encoding }
\label{ex:rel-prov}
\end{table}

\subsection{Proof of Correctness [Sketch]}\label{sec:correctness}
\rev{We will prove that \FMain is a lossless compression algorithm. The basic proof strategy is to show that every step of this algorithm is lossless and fully reversible, and thus, by induction, the algorithm is also lossless.}

\rev{\noindent \textbf{Union of Cartesian Products. } Let $R$ be a relational instance with set semantics over the attributes $Y$. Every such instance can be expressed as a ``union of Cartesian products'' (UCP). For example consider the set of 2-tuples $\{('a',1), ('a', 2), ('b',3)\}$, one can express this set as $\{ \{'a'\} \times \{1,2\} \cup \{'b'\} \times \{3\}\}$. The Cartesian product generates the columns, and the union combines the rows. This representation is identical to the $\mathcal{T}$ representation in the worst-case optimal join literature ~\cite{ciucanu2015worst}. This representation is fully lossless where: 
\[\{('a',1), ('a', 2), ('b',3)\} \equiv \{ \{'a'\} \times \{1,2\} \cup \{'b'\} \times \{3\}\}.\]
Note that ``set semantics'' is important here; this approach is lossless when every row in the lineage relation is unique (which is true in \sys).} 

\rev{\noindent \textbf{Multi-Attribute Range Encoding as a UCP.} Every step of Multi-Attribute Range Encoding produces a valid UCP representation of the instance. For example, table \ref{ex:input-prov} shows the algorithm's output for an example table. In UCP notation, 
\[
\{ \{1\} \times \{1\} \times \{1,2\} \cup \{2\} \times \{2\} \times \{1,2\} \cup \{3\} \times \{3\} \times \{1,2\} \}\]}
\rev{This is because we only replace contiguous ranges where ALL other attributes match on equality.}

\rev{\noindent \textbf{Relative Transformation. } The results above hold for any integer-valued dataset. Thus, we can always choose how to encode these integers as long as that encoding is reversible. The relative transformation simply encodes certain columns as a delta rather than an absolute number (output-to-input).} 

\subsection{Asymmetry in Representation} 
Our lineage representation is asymmetrical - input attributes can have relative indices, but output attributes are absolute. While not relevant for correctness (lossless compression), it does have an impact on query processing. We can push down predicates only on absolutely indexed columns.
This is easily resolved by materializing a different version of the compressed tables with the inverse property when necessary - a version where output attributes can have relative indices, but input attributes are absolute. Either both versions can be stored on one version depending on the distribution of forward and reverse queries.

In Table \ref{ex:rel-forward}, we show the forward query table of our running example. This table encodes the same information as the previous representation but has absolute values on attributes $a_1,\dots, a_m$.
\begin{table}[t]
\footnotesize

\centering
\begin{tabular}{lllll}
\hline
$a_1$ & $a_2$ & $b_1$ & $b_1a_1$ & $b_1a_2$ \\\hline
[1,3] & [1,2] & - & 0 & - \\
\end{tabular}
\caption{Alternative Compression Representation to Enable Forward Queries }
\label{ex:rel-forward}
\end{table}

\section{In Situ Query Processing} \label{sec:queries}
We now describe the algorithm for directly answering backward and forward queries over the compressed lineage index. 

Recall the user calls the method \texttt{prov\_query} to query the lineage between two Array along some path represented by chained Arrays $X^1 \xrightarrow{} X^2 \dots \xrightarrow{} X^n$. This query returns the cells in $X^n$ that contribute to particular cells in $X^1$. The user specifies the interested cells with $Q$, a table with attributes $\Omega_{X^1}$ (which correspond to array indices of $X^1$) that contain the interested cell indices. Such a query is answerable if a chain of operations connects the given path, and the lineage of these operations is stored in \sys. 

\subsection{Basic Query Processing}
Suppose the uncompressed lineage relation between array $X^i$ and $X^{i+1}$ is stored as a full lineage table $R^i$. To materialize the lineage relationship along the chain of operations, we join these tables with a natural join:

\[
R^1 \bowtie ... \bowtie R^{n-1}
\]

In forward and backward queries, the query is simply a predicate over $R^1$, specifically over the attributes corresponding to $X^1$. This can be elegantly expressed as another join: 

\[
Q \bowtie R^1 \bowtie ... \bowtie R^{n-1}
\]

\subsection{In-Situ Join Algorithm}
Instead of working with the original relations, we now work with compressed relations. We assume that each $R_1^i$ table is appropriately constructed from the full lineage table $R^i$ using the \FMain compression algorithm. Now, we can perform a similar query over the compressed tables \emph{but replace the equality joins with custom $\theta$-joins over intervals}. These $\theta$-joins account for the range-encoding and relative value transformations of \FMain.  

\[
Q' \bowtie_{\theta} R_1^1 \bowtie_{\theta} ... \bowtie_{\theta}R_1^{n-1}
\]

The query, $Q'$, is encoded from $Q$ in the same format as the compressed relational lineage tables with multi-attribute range encoding. This encoding is typical with other in-situ query processing algorithms, e.g., dictionary coding~\cite{abadi2006integrating}.

There are two subroutines within a $\theta$-join:
\begin{itemize}
    \item Range Join. Perform a range join on the range intervals in the compressed tables \cite{range_join}.  \rev{Recall, after compression, values in the table can be ranges. We want to join on ranges that have any overlap. This is simply a range join, where the join condition is $range_1 \cap range_2 \ne \emptyset$.}
    \item De-Relativize Indices. Convert the relative attributes of the resulting table into absolute attributes.
\end{itemize}

We will now describe the $\theta$-join over tables $Q'$ and $R_1^1$:

\[
Q' \bowtie_\theta R^1_1(X^1, X^2)
\]

Using our running example, we also demonstrate how a query is processed with a $\theta$-join. Assume that Table \ref{ex:rel-prov}, $R_{exp}$ is stored in our system. Let us generate a query over the path $B \to A$ (i.e., a backward query) with the cell specification table, $Q_{exp}$, as shown in Table \ref{tab:query_q}. This specifies that we are interested in the cells with indices $b_1 = 1, 2$.

\begin{table}[ht!]
\footnotesize
\centering
\begin{tabular}{lllll}
\hline
$b_1$  \\\hline
[1,2] \\
\end{tabular}
\caption{Example Query for Stored Lineage Table}
\label{tab:query_q}
\end{table}

\subsubsection{Step 1. Range Join} 

As the first step of the $\theta$-join, a range join is performed over $Q$ and $R^1$. \rev{Given columns that describe intervals in two tables, a range join joins on the intersection between those intervals. For example, we have a range join over $\alpha$ and $\beta$ columns as:}

\begin{table}[ht!]
\rev{
\begin{tabular}{l}
\hline
$\alpha$ \\\hline
$[1, 3]$ \\
$[4, 8]$ \\
\end{tabular}
}
\rev{
$\bowtie_{\text{Range}}$
}
\rev{
\begin{tabular}{l}
\hline
$\beta$  \\\hline
$[0, 2]$ \\
$[10, 12]$ \\
\end{tabular}
}
\rev{
$\rightarrow$
}
\rev{
\begin{tabular}{lll}
\hline
$\alpha$ &$\beta$ & $\alpha\beta\text{(Intersect)}$  \\\hline
$[1, 3]$ & $[0, 2]$ & $[1, 2]$ \\
\end{tabular}
}
\end{table}

In our case, this join takes the \rev{joint} intersection of ranges over all common attributes of $\Omega_{X^1}$ between the two tables. We now use the critical insight that each row in $R^1$ describes an all-to-all relationship between input and output attributes. This is true even in the case of relative attributes; however, in this case, the all-to-all relationship is in the relative space. Using this insight, we can observe that a range join \rev{between $Q$ and $R^1$} preserves the lineage relationships described in $R^1$ for the relevant cells given by $Q$.

\begin{figure}
\tiny
  \centering
  \begin{tabular}{c}
{\textbf{(A) Compressed lineage of an All-to-All Operation}}\\ \\
\includegraphics[width=.25 \linewidth]{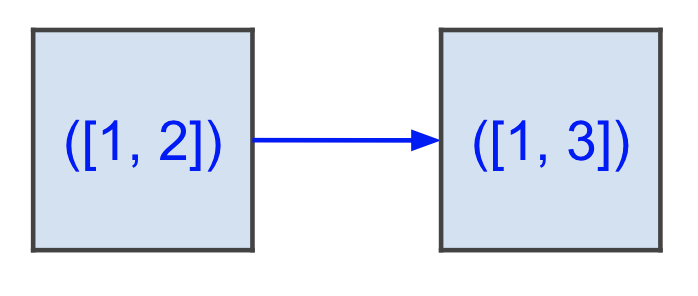}
\end{tabular} 

\begin{tabular}{c}
{\textbf{(B) Intersection with [1,2] Query}}\\ \\
\includegraphics[width=.25 \linewidth]{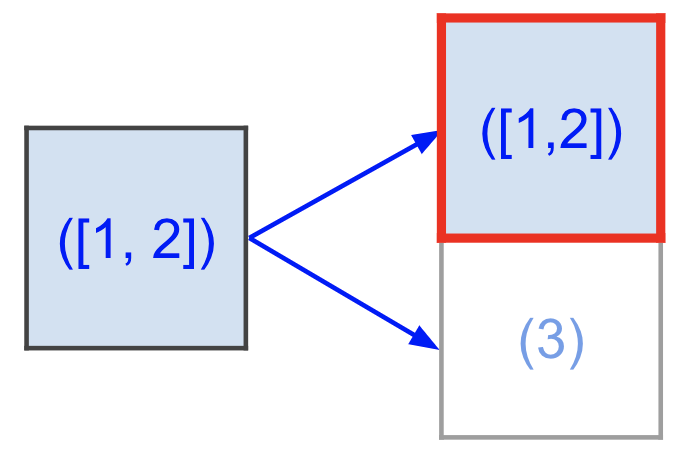}
\end{tabular}
\begin{tabular}{c}
{\textbf{(C) Get Full Range of First Array}}\\ \\
\includegraphics[width=.25 \linewidth]{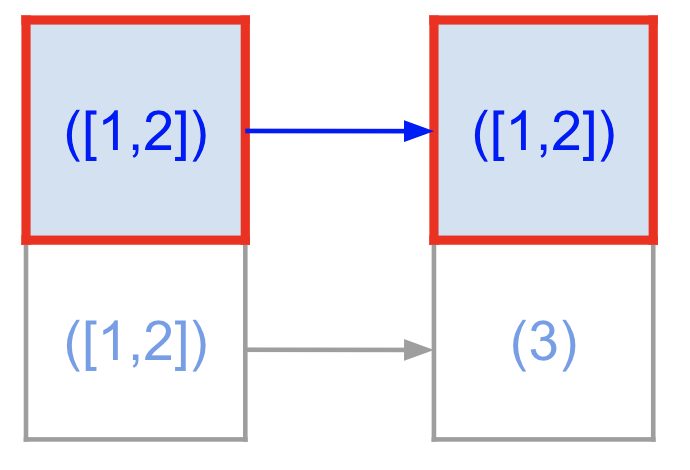}

\end{tabular}
    \caption{This is an example of a range join preserves lineage, where, given a lineage table compressed with \protect\FMain (A), we can identify the intersections between the query and the table (B) and then get the full lineage of those intersecting intervals (C). }
    \label{example:join}
    
\end{figure}

Figure \ref{example:join} demonstrates an example of this insight, given a compressed array lineage relationship between array 
attribute intervals: $[1,2] \rightarrow [1,3]$ (A). Suppose we query for the backward lineage of the second array at indices (1,2). We are only interested in the lineage relationships described by the intersection between the query and the lineage table (on the second array's attributes) (B). Due to the all-to-all relationship, we want the full range of the first array's attributes corresponding to those intervals (C). This is elegantly represented with a range join. 

After performing the range join over tables $Q_{exp}$ and $R_{exp}$ from our running example, we get the table shown in Table \ref{tab:query1}. Note that, \rev{over the intersection of ranges}, the lineage relationships captured by full lineage table $R_{exp}$ are preserved, but only over the cells specified by the query $Q_{exp}$.

\begin{table}[ht!]
\footnotesize
\centering
\begin{tabular}{llllll}
\hline
$b_1$ \rev{(Intersect)} &$a_1$ & $a_1b_1$ & $a_2$ & $a_2b_1$ \\\hline
[1,2] & - & 0 & [1, 2] & - \\
\end{tabular}
\caption{Range Joins Preserve the Lineage Relationships of Queried Cells}
\label{tab:query1}
\end{table}

\subsubsection{Step 2. De-Relativize Attributes} 

Let the output table of the range join over $Q$ and $R^1$ be represented as $T$. The table $T$ may have relative attributes; however, we would like the $\theta$-join operation to return a table with only absolute attributes for future $\theta$-joins or for displaying the query result. As the second step of the $\theta$-join, we shall now demonstrate how to convert the relative attributes in $T$ back into absolute attributes. 

\begin{figure}
\tiny
  \centering
  \begin{tabular}{c}
{\textbf{(A) Compressed Lineage of an One-to-One Operation}}\\ \\
\includegraphics[width=.25 \linewidth]{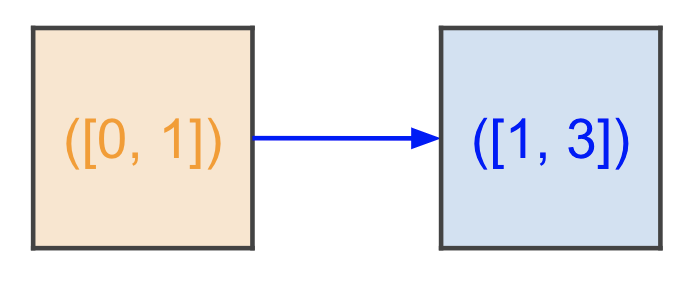}
\end{tabular} 

\begin{tabular}{c}
{\textbf{(B) Range Join with [1,2] Query}}\\ \\
\includegraphics[width=.25 \linewidth]{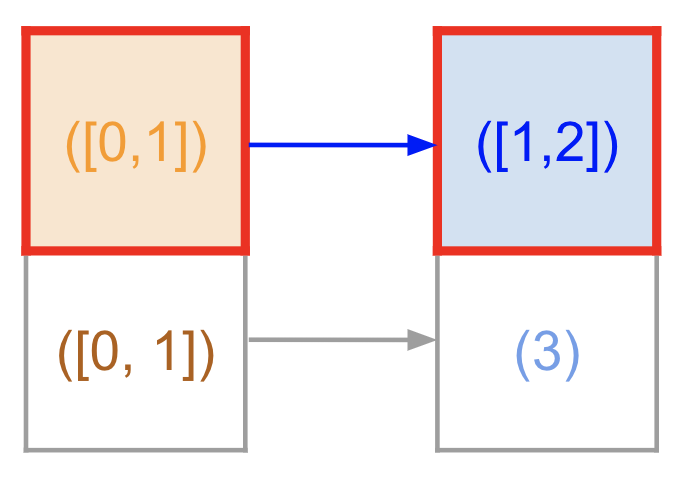}
\end{tabular}
\begin{tabular}{c}
{\textbf{(C) Convert into Absolute Indices}}\\ \\
\includegraphics[width=.25 \linewidth]{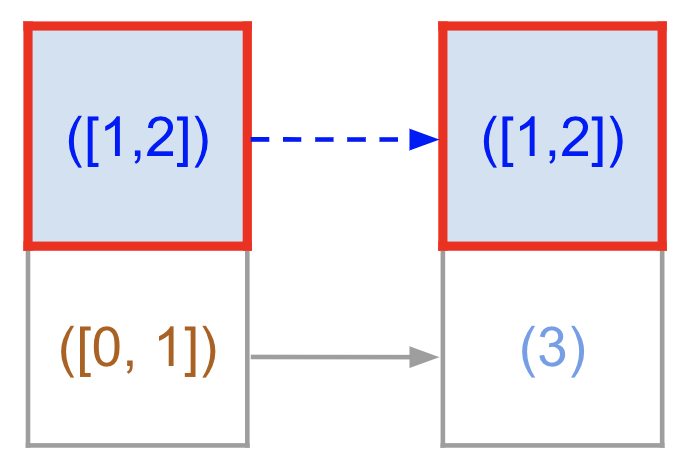}

\end{tabular}
    \caption{We have an example of how we can find absolute indices from relative range intervals, where, given a lineage table compressed with \protect\FMain (A), we can perform a range join with a query (B), and then calculate the absolute indices from the resulting table (while breaking the all-to-all relationship) (C).}
 \label{example:derelative}
    
\end{figure}

The relative attributes can be combined with the absolute attributes they reference to generate the original absolute values. We first give an intuition for this process with the example shown in Figure \ref{example:derelative}. This example shows a lineage relationship between two array index intervals where the first index is \emph{relative}: $[0,1] \rightarrow [1,3]$ (A). Given a query for the backward lineage of the second array at index (1,2), we can calculate the relevant lineage relationships with a range join (B). Now, we can calculate the absolute indices captured in these relationships by reversing the relative operation (C). This breaks the all-to-all lineage relationship (as shown with a dashed arrow) but still gives the correct lineage cells for the queried intervals. Note that this calculation is made without expanding any intervals into full representations, and \emph{the absolute interval values remain in compressed range encoding}. \rev{By working through the transformations of \FMain, it can be shown that for any arbitrary ranges $t.x, r.x, t.xy$, the original absolute indices captured by the ranges are exactly contained in the output of $rel\_for$ or $rel\_back$.}

We can generalize this process to create a reversal algorithm for any row $t \in T$. To formalize this, for attribute $y \in \Omega_{X^2}$, we perform the following algorithm:

\begin{itemize}
      \item If $t.y$ is not empty, do nothing.
      \item Else, find a non-empty $t.xy$ for some $x \in \Omega_{X^1}$, 
      \begin{itemize}
      \item If the table is a backward representation, \\ set $t.y := rel\_back(t.x, t.xy)$
      \item If the table is a forward representation, \\ set $t.y := rel\_for(t.x, r.x, t.xy)$
      \item Set $t.xy := \emptyset$
      \end{itemize}
\end{itemize}

We compute the absolute interval values using the methods $rel_{for}$ and $rel_{back}$, for forward and backward tables, respectively:

\[rel\_back(t.x, t.xy) := [t.x[1] + t.xy[1], t.x[2] + t.xy[2] ]\]
\[rel\_for(t.x, r.x, t.xy) := \]
\[[\max(t.x[1] - t.xy[2], r.x[1] - t.xy[2]), \]
\[\min(t.x[2] - t.xy[1], r.x[1] - t.xy[2])]\]

For forward queries, the method $rel\_for$ also needs the values of row $r \in R^1_1$ from which $t$ is derived.

In the running example, we can apply this calculation to the single row in Table \ref{tab:query1} to retrieve only absolute attributes. This final table is shown in Table \ref{tab:query_abs}. This final table shows the cells in $A$ that contributed to the cells in $B$ with indices $b_1 = 1, 2$. Since the example query only has one step in the lineage path, \texttt{query\_cells} returns $(a_1: [1,2], a_2: [1, 2])$. For a general query, we repeat the $\theta$-join for $T \bowtie_{\theta} R^2(X^2, X^3)$ and so on. \rev{Each join returns the relevant indices of the following array in the path until we reach the final array.}

\begin{table}[ht!]
\footnotesize
\centering
\begin{tabular}{lll}
\hline
$b_1$ & $a_1$ & $a_2$ \\\hline
[1,2] & [1,2] &[1, 2] \\
\end{tabular}
\caption{Final Table Contains Only Absolute Values}
\label{tab:query_abs}
\end{table}

\subsubsection{Query Plan and Optimizations} We process these joins in a left-to-right query plan, i.e., we process the joins in the order that they appear in the path specified by the user. To optimize performance, \sys apply projection and row-reduction operations on the output of the $\theta$-join, $T$, before the next join. First, \sys finds the projection $T': \Pi_{\{y \in \Omega_{X_2} \}}(T)$ (since the query now only requires $X_2$ attributes). Second, any rows in $T'$ describing cell indices with adjacent intervals are merged into a singular row, using an algorithm similar to the multi-attribute range encoding step in \FMain. This minimizes the total number of rows in $T'$, reducing the cost of the next join.
\section{Lineage Reuse}\label{sec:reuse}

When the same operations are repeatedly called on different input arrays, the lineage from one call can populate the lineage from another. For example, if the same featurization function is used on a training dataset and then a testing dataset, we may find the lineage of that function on the testing dataset without explicit capture. 

This section describes how previously captured lineage can be reused during repeated operation calls in \sys. Our core contribution here is allowing the reuse of lineage from operations with differently shaped array inputs through ``index reshaping''. This reshaping is explicitly enabled by extending our compression algorithm, \FMain. 

\subsection{Basic Design}
An operation signature is associated with each call of \texttt{register\_operation}. The operation signature is defined by the name of the array operation, \texttt{op\_name}, the set of input arrays, \texttt{in\_arrs}, and other input arguments, \texttt{op\_args}. Users specify the relevant input arguments in \texttt{op\_args}. We require that they define the lineage pattern of the operation for given array inputs up to pseudo-randomness.

\sys can internally map this operation signature with the associated lineage tables of its input and output arrays:

\begin{lstlisting}
base_sig(op_name: String, in_arrs: String [], op_args: Any[]) = {R}
\end{lstlisting}

In subsequent calls to \texttt{register\_operation} with the same signature, calls to the capture object can be bypassed by the previous lineage tables. This basic strategy has been employed in systems like Lima~\cite{phani2021lima}. However, it requires recording and matching on the entire input array and generally has limited reuse capabilities. As we will see in the following section, our system goes a few steps further regarding reuse by defining more general signature mappings.

\subsection{Lineage Extrapolation}
In several scenarios, the lineage depends only on the shape of the input arrays and not the actual input arrays themselves. For example, the lineage for linear algebra operations or a forward evaluation of a neural network on a different mini-batch of data fits this criterion.

In those cases, we can store a more general mapping between operation signatures and the lineage table: 

\begin{lstlisting}
dim_sig(op_name:String, in_shapes:Tuple[], op_args:Any[]) = {R}
\end{lstlisting}

, where \texttt{in\_shapes} is the shape of arrays in \texttt{in\_arrs} of the operation signature. Similar to the previous case, in subsequent calls to \texttt{register\_operation}, we can directly reuse the lineage registered above if there is a match on the operation signature. However, a match only needs the same input array shapes rather than the same input arrays.

Finally, it is also possible to extrapolate lineage in some scenarios without matching on shape. For example, consider an image smoothing operation with a convolutional filter. The user runs this program twice, and for the second time, the image is down-sampled beforehand. In this case, the lineage relation from the second run is simply a fraction of the relation of the first run. We have found that this pattern is relatively common in data science workflows.

In \sys, we can generate the lineage table to any input array shape for this lineage pattern by ``index reshaping'' the original compressed lineage table into a generalized representation. Now, \sys can map operation signatures to lineage tables without requiring any matches on the input array shapes:

\begin{lstlisting}
gen_sig(op_name:String, op_args:Any[]) = {R'}
\end{lstlisting}

, where $R'$ is a modified table that enables index reshaping. We can generate $R'$ directly because of \sys's compressed representation of lineage tables. 

To create $R'$ given a compressed relational table, $R$, for $d_i \in dim(X)$, we identify all intervals equal to $[1,d_i]$ in $R$ and replace them with the generalized interval $[1, D_i]$, where $D_i$ is an indicator of that attribute. Suppose those intervals are the only ones dependent on the shape of the input arrays, and there is no data dependence. In that case, a generalized representation is formed independent of the input array shape. The symbolic interval is used to extrapolate the lineage for calls of the same function but with differently shaped data.

\begin{figure}
\tiny
  \centering
  \begin{tabular}{p{0.4\linewidth}}
{\textbf{(A) Compressed Lineage of an Aggregate Operation for a 1-D Array with $d_1 = 2$}}\\ \\
\includegraphics[width=0.9\linewidth]{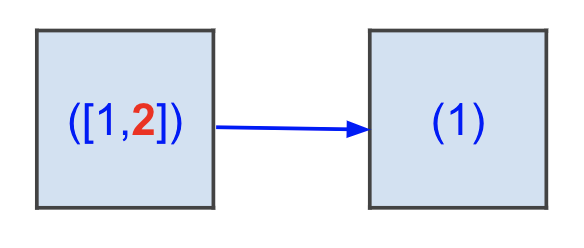}
\end{tabular}
\begin{tabular}{p{0.4\linewidth}}
{\textbf{(B) Generalize Representation of Lineage from (A)}}\\ \\
\includegraphics[width= 0.9\linewidth]{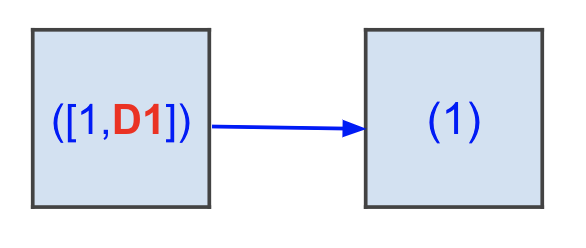}
\end{tabular}

\begin{tabular}{c}
{\textbf{(C) Converted Lineage from (B) for a 1-D Array with $d_1 = 4$}}\\ \\
\includegraphics[width=.3 \linewidth]{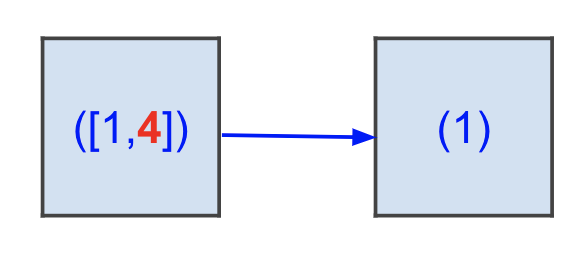}
\end{tabular}
\caption{This figure shows how compressed lineage tables can be leveraged for reuse with index reshaping. The compressed lineage of an aggregate operation for a two-cell array (A) can be converted to a generalized representation (B), which can be used to find lineage for downstream operation calls with different array shapes (C).}
\label{example:general}
\end{figure}

Figure \ref{example:general} shows an example of index reshaping. Given an aggregation operation that captures an all-to-all relationship and outputs a 1-cell array, and a 1-D input array with $d_1 = 2$, \sys can generate a compressed lineage table with \FMain (A). For the generalized representation, the interval in this compressed lineage is converted to $[1, D_i]$ based on the algorithm above (B). In a subsequent call to the same aggregation operation with an input array with $d_1 = 4$, \sys can set $D_1 = 4$ to capture that call's compressed lineage without additional tracking (C).

\subsection{Automatic Reuse Prediction} 

Understanding when reuse is possible requires knowledge of the exact lineage pattern of an operation and how that lineage pattern is stored. This can be time-intensive and error-prone for end-users. Instead of requiring users to enable reuse, \sys can try to automatically predict when \texttt{dim\_sig} and \texttt{gen\_sig} signatures can be generated. This is based on simple heuristics that extrapolate lineage based on the past behavior of that operation signature.

When \texttt{register\_operation} is called, \sys internally stores \emph{temporary} \texttt{dim\_sig} and \texttt{gen\_sig} mappings. In the subsequent $m$ calls to \texttt{register\_operation} that match on either partial operation signature, \sys checks if the stored lineage tables in these mappings match the new lineage tables. We set the mapping relations as permanent if a match occurs after $m$ calls. Otherwise, we mark that partial signature key as not reusable. For the \texttt{gen\_sig} matches, we also require that the $m$ calls have different array shape values. In our current implementation, we set $m = 1$.

\section{Experiments}\label{sec:experiments}
 We performed experiments to evaluate each component of \sys. Our performance experiments were performed on Chameleon, a large-scale computer science research platform~\cite{cloud2017configurable}, with an Intel Xeon Gold 6126 Processor, an RTX6000 GPU, and 192 GiB of RAM.

\subsection{Prototype Capture Algorithms}
In \sys, we include several prototype capture methods for different types of operations in the Python data science ecosystem. These are included to demonstrate our system but are not meant to represent what is required to capture lineage for entire array workflows. We view the difficult task of lineage capture in data science as perpendicular work to this paper.

\subsubsection{Cell-Level Tracking in \texttt{numpy}} \sloppy The \texttt{numpy} array is a popular array framework in Python. Using annotations, we have implemented a data type \texttt{tracked\_cell} that automatically tracks the lineage of every cell within a \texttt{numpy} array during low-level operations. Under the hood, this capture technique is similar in semantics to taint tracking~\cite{zhu2011tainteraser}. With efficient memory management, \texttt{tracked\_cell} performs 300x over the equivalent Python implementation and can scale to (and over) arrays with 1 billion cells. The lineage history of each cell in the output array can be used as the \texttt{capture} argument in a \texttt{Lineage} object. Detailed capture performance experiments and further descriptions will be included in the Appendix of the technical report.

\subsubsection{Explainable AI Tracking}
\sys also allows lineage capture with lineage derived from model explanation algorithms. We assume that the model operations take the form $y = \texttt{model}(x)$, where $x$ is an input example a (vector, 1-d array) and y is the predicted label (a vector, 1-d array). Under this assumption, \sys can apply explainable AI techniques - i.e., LIME or D-RISE - over the model ~\cite{ribeiro2016should, Petsiuk2021drise}. Both explainable AI techniques generate a bipartite weighted contribution relationship between the cells in $x$ and the cells in $y$. Direct contributions are filtered from this graph based on a significance threshold.

\subsubsection{Custom Tracking Functions} Finally, we have implemented custom `group-by' and  `inner-join' operations that record the lineage history of individual cells upon execution. This demonstrates the integration of \sys with traditional relational operations.

\subsection{DPSM Baselines and Implementation}
We have implemented alternative storage formats and physical designs to evaluate the compression and query components of \sys. \rev{For most baselines, the query interface is served by the DuckDB system~\cite{raasveldt2019duckdb}. We additionally evaluated one array-based baseline, where the query is completed with array operations}. All storage formats \rev{(except the array-based one)} are designed to be transparently queried from DuckDB; however, some baselines might require explicit decompression before querying.


\textbf{Raw.} The captured lineage tuples are stored in a row-oriented format without data compression. This format is similar to the design principles of Ground~\cite{hellerstein2017ground}, but with an implementation in DuckDB so we could compare to baselines~\footnote{We referred to the table design in \textbf{https://github.com/ground-context/ground}}. 

\rev{\textbf{Array.} The captured lineage tuples are stored as a numpy array.}
    
\textbf{Parquet.} The captured lineage tuples are stored directly in Apache Parquet files, a columnar file format that different query processing engines can parse. We use default encoding and row-group partitioning settings.
    
\textbf{Parquet-GZip.} The captured lineage tuples are stored in a GZIP compressed Parquet file, as suggested by industry practice~\cite{pqtbest}.
    
\textbf{Turbo-RC.} We designed a custom columnar format that applies state-of-the-art integer compression over each column. This compression included run-length encoding combined with integer entropy coding~\cite{hanzo2002turbo}. 

We compared these baselines to \FMain and \FMain-GZip, a version of \FMain where GZip is applied to the results of the main algorithm, similar to Parquet-GZip. The GZip compression had minimal performance impact on our system, with significant storage benefits for unstructured lineage. In practice, \sys uses the \FMain-GZip algorithm for optimal storage across all operations.

\begin{table*}[htbp]
\footnotesize
\centering
\begin{tabularx}{2\columnwidth}{l|llllllllllllllll}
\hline
\multicolumn{1}{c|}{ \textbf{Name}} & \multicolumn{1}{c|}{\textbf{Raw}} & \multicolumn{2}{c|}{\rev{\textbf{Array}}} & \multicolumn{2}{c|}{\textbf{Parquet}} & \multicolumn{2}{c|}{\textbf{Parquet-GZip}} & \multicolumn{2}{c|}{\textbf{Turbo-RC}} & \multicolumn{2}{c|}{\textbf{\FMain}} & \multicolumn{2}{c}{\textbf{\FMain - GZip}} \\
\hline
 & Abs & Abs & Rel & Abs & Rel & Abs & Rel & Abs & Rel & Abs & Rel & Abs & Rel  \\
&  (MB) & \rev{(MB)} &  \rev{(\%)} &  (MB) &  (\%) &  (MB) &  (\%) &  (MB) &  (\%) &  (MB) &  (\%) &  (MB) &  (\%) \\
\hline
 Negative   & 22.66 & \rev{32.00}&\rev{141} & 5.05 & 22.31 & 4.33 & 19.10 & 5.11 & 22.58 & \textbf{0.00978} & \textbf{0.0431} & 0.0103 & 0.0457 \\

 Addition   & 45.33 &\rev{64.00}& \rev{141}&  10.11 &  22.31 &  8.66 & 19.10 & 10.23  &  22.58 &  \textbf{0.0195} &  \textbf{0.0431}  & 0.0207 & 0.0457 \\
 
Aggregate  & 20.55 &\rev{32.00}& \rev{155}&  0.131 &  0.639 &  0.0255 & 0.124  & 3.73  &  18.17 &  \textbf{0.00978} &  \textbf{0.0475} & 0.0103 & 0.0504 \\
   
 Repetition & 98.22 &\rev{128.0}&\rev{130} &  25.20 &  25.65 &  14.52 & 14.78  & 22.03  &  22.43 &  \textbf{0.00983} & \textbf{0.0100} & 0.0103  & 0.0105 \\
   
 Matrix*Vector   & 39.22 &\rev{64.00}& \rev{163}& 0.254 &  0.649  &  0.048 & 0.122 & 6.77 &  17.25 &  \textbf{0.0195} &  \textbf{0.0498} &  0.0207 & 0.0528  \\

Matrix*Matrix   & \rev{40127} &\rev{64000} & \rev{159}& \rev{254.8} &  \rev{0.635}  & \rev{48.06} & \rev{48.06} & \rev{0.119} &  \rev{6770.} &  \rev{\textbf{0.0198}} &  \rev{\textbf{4.95}$\times 10^{-5}$ } & \rev{\textbf{0.0210}} &  \rev{\textbf{5.23}$\times 10^{-5}$ }   \\

Sort      & 22.66  &\rev{32.00}&\rev{141} & 3.38 & 14.92 & \textbf{2.76} & \textbf{12.19} & 6.10 &  26.91 &  3.43 & 15.15 & 2.79 & 12.33   \\
   
ImgFilter & 4.288 &\rev{5.632}&\rev{131} & 1.96 & 45.93 &  1.06 & 24.73 & 1.05 &  24.64 &  \textbf{0.00999} &  \textbf{0.233} & 0.0104 & 0.244  \\
   
   \hline
Lime      & 24.09 &\rev{29.63}& \rev{123}& 0.529 & 2.19 &  0.123  & 0.513  & 5.97 &  24.78 &  0.0123 &  0.0511 & \textbf{0.0121} & \textbf{0.0502}  \\
   
 DRISE     & 8.82 &\rev{11.07}&\rev{125}& 0.0898 & 1.01 &  8.242 & 0.271  & 2.19 & 24.91 &  \textbf{0.0106} &  \textbf{0.120} & 0.0108 & 0.123 \\
   \hline
Group By & \rev{183.43} &\rev{251.00}&\rev{136} &\rev{32.62} & \rev{17.78} & \rev{\textbf{13.56}} & \rev{\textbf{7.39}} & \rev{35.97}  & \rev{19.61}& \rev{29.45} &  \rev{16.05}  & \rev{13.62} & \rev{7.42} \\
Inner Join & \rev{2330} &\rev{2612.13}& \rev{111}& \rev{195.44} & \rev{8.36} &  \rev{53.44} & \rev{2.28} & \rev{584.98} &  \rev{25.02} & \rev{14.13} &  \rev{0.604} & \rev{\textbf{6.36}} & \rev{\textbf{0.272}} \\
\end{tabularx}

\caption{Comparison of Compression Ratio for Different Algorithms Against \protect\FMain } \label{tab:compression_result}
\end{table*}

\subsection{Compression} \label{sec:exp-compression}
\subsubsection{Lineage Capture Storage Size}
We compare the size on disk of the different compression formats for tables within our lineage index. Each baseline, \FMain and \FMain-GZip, were applied to compress the lineage of selected operations.

Twelve individual data science operations were constructed to capture a full range of individual array operations. This workload can be broken down into the following categories:


\begin{enumerate}
    \item \rev{Eight} of those operations are \texttt{numpy} operations \cite{extended_experiments}: for five of the operations (Negative, Addition, Aggregate, Repetition, Matrix*Vector, Matrix*Matrix), the lineage is independent of the data values, and for two of the operations, the lineage is value-dependent (Sort, ImgFiter). The lineages are tracked with low-level array tracking. Note that ``Sort'' is the worst case for \FMain, where no continuous patterns exist in the lineage. 
    \item Two of these operations are explainable AI. A frame was chosen from VIRAT~\cite{oh2011large}, a surveillance camera dataset, and YOLOv4 object detection was applied over that frame to detect a 'car' object. On this model and data, lineage was captured with LIME and DRISE.
    \item  The final two operations are relational operations, and the stored lineage is captured with custom functions \cite{cheney2009provenance}. These operations are applied to \rev{the full IMDB dataset tables}\cite{IMDb_noncommercial_datasets}. Note that the 'tconst' and 'startYear' columns are sorted in the original tables, but 'isAdult' is unsorted.
\end{enumerate}

Table \ref{tab:compression_result} shows the size of the lineage file on disk for each compression algorithm and relative compression ratios to the raw file. \rev{The Array baseline is uncompressed and has a similar storage size to Raw}. Across all operations, Turbo-RC has the most consistent performance - suggesting that its compression algorithm did not exploit specific underlying lineage patterns. We can observe that Parquet and Parquet-GZip are exceptionally efficient in compressing aggregation patterns - compressing significantly on ``Aggregate'' and ``Matrix*Vector '' - but are still less optimal in those operations than \FMain and \FMain-GZip. 

Over operations that explicitly matched the patterns described in Section \ref{sec:compression} (``Negative'', ``Addition'', ``Aggregate'', ``Repetition'', ``Matrix*Vector'', \rev{``Matrix*Matrix''}), \FMain is the most efficient compression algorithm, achieving a compression ratio of $5 \times 10^{-5}\%$ and beating the closest baseline by up to \rev{2000x}. Even over operations with only partially structured lineage (``ImgFilter'', ``Lime'',  ``DRISE'', and ``Inner Join''), \FMain performs exceptionally well, with a compression ratio of $<0.3\%$, beating the closest baselines by up to 400x. We attribute this result to the algorithm's ability to compress more complex multi-columnar patterns and efficiently represent range intervals. 

\FMain-GZip performs slightly worse than \FMain over the most structured lineages due to the slight overhead of the GZip compression. However, for partially structured lineage, it was better than \FMain in two cases (`Lime'' and ``Inner Join''). The benefit of \FMain-GZip is over unstructured lineage, significantly improving the lineage storage size of two unstructured operations (``Group-By'', ''Sort). In both cases, it was within 0.5\% of the optimal compression algorithm. These experiments show that applying \FMain-GZip to array lineage tables captured within \sys is feasible and generally the most optimal solution.

We only materialize the \FMain and \FMain-GZip representation for backward queries. Since the backward and forward queries store the same information, only one must be stored in long-term storage. We measured the file size of the database files that were ultimately served to DuckDB.

\subsubsection{Compression Latency}
Next, we evaluated the compression latency of the different baselines and \FMain-GZip. This latency considers the entire read, format-conversion, compression, and flush latency time it takes to write the compression lineage tables to disk. We measured the latency of table compression for the two extreme lineage types: one-to-one element-wise operations and one-axis aggregation operations. This latency is measured over a range of array sizes.

\begin{figure}
  \centering
  \includegraphics[width=.46\linewidth]{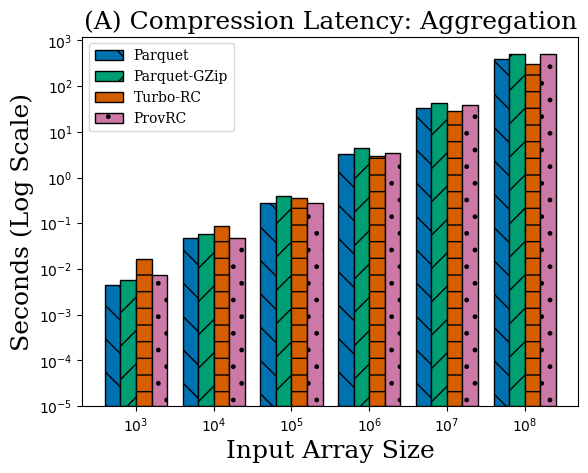}
  \includegraphics[width=.46\linewidth]{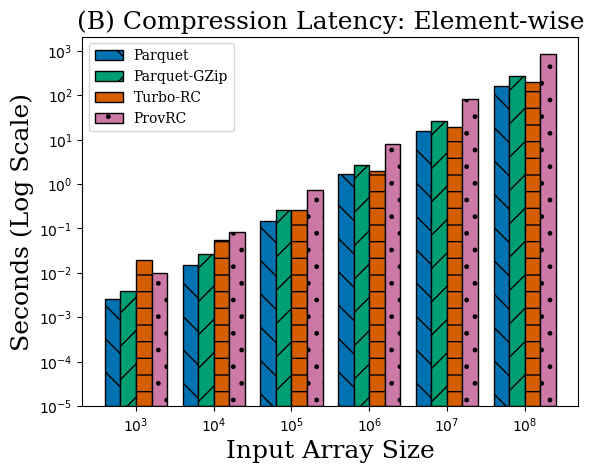}
    \caption{(A) and (B) benchmark the compression algorithms in terms of their latency as a function of input size.}
    \label{exp:compperf}
\end{figure}

Figure \ref{exp:compperf} shows the latency of the different compression algorithms. We note that the latency between different compression algorithms is within an order of magnitude of each other and increases as the input array shape increases. \FMain-GZip is very efficient on smaller tables for aggregation patterns; however, it is consistently less efficient in other cases and is relatively slow in compressing larger element-wise lineage tables compared to baselines. Given that the current implementation is written in Python for simplicity, and the baselines are written in C++ and are from mature/public codebases, we find this performance reasonable. \FMain is also highly parallelizable, so we expect significant performance gains from a multi-threaded implementation. 

\subsection{Query Processing} \label{sec:exp-query}

\begin{figure}
  \centering
  \includegraphics[width=.49\linewidth]{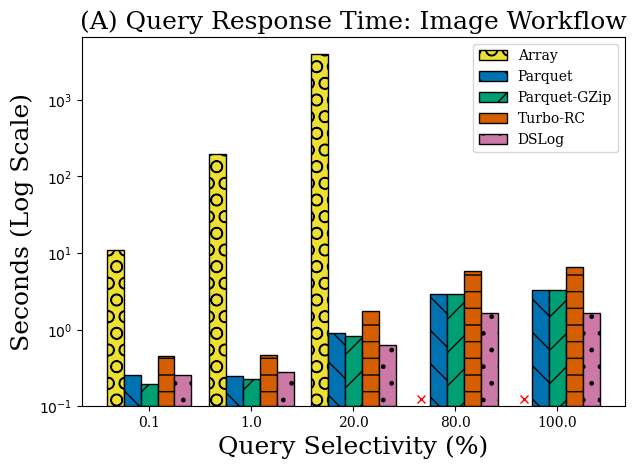}
  \includegraphics[width=.49\linewidth]{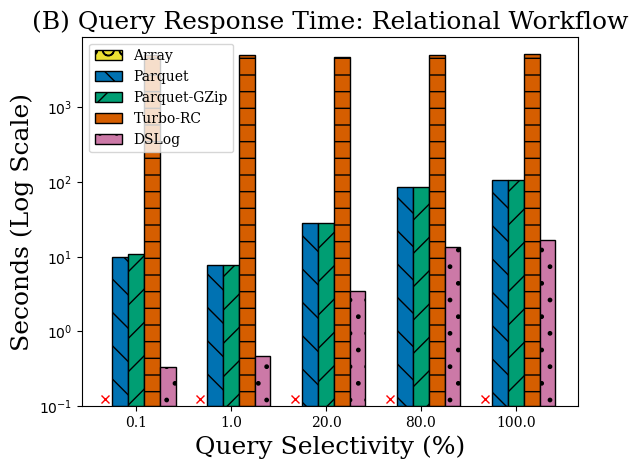}
   \includegraphics[width=.49\linewidth]{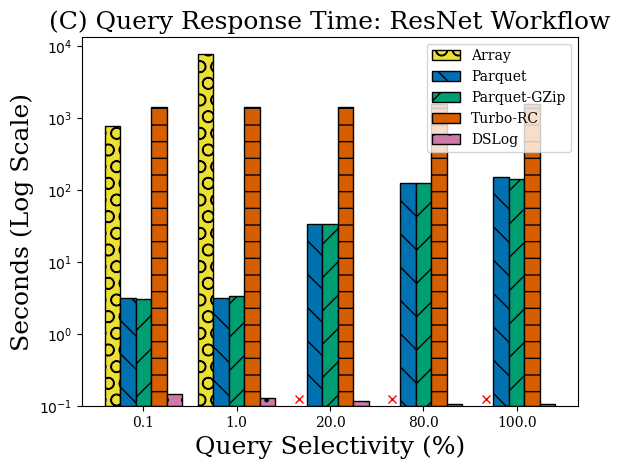}
    \caption{\rev{These graphs show the evaluation of query latency on (A) image machine learning, (B) relational and (C) ResNet workflows.}}
    \label{exp:custom_q}
\end{figure}

\begin{figure}
  \centering
  \includegraphics[width=.8\linewidth]{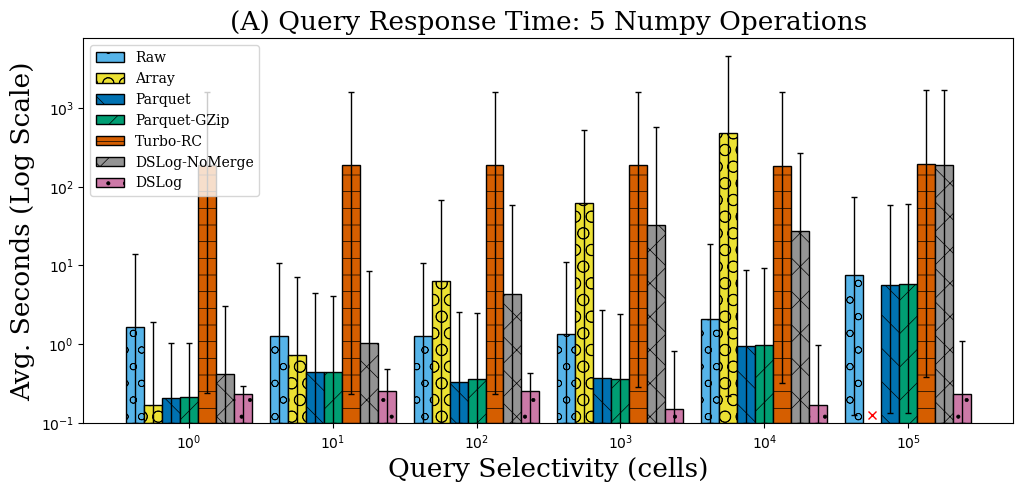}
  \includegraphics[width=.8\linewidth]{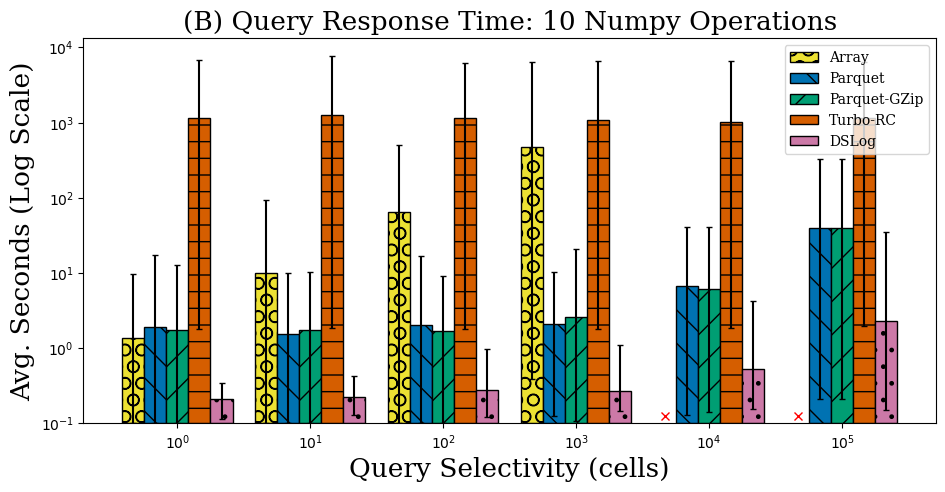}
    \caption{\rev{These graphs show the evaluation of average query latency on random \protect\texttt{numpy} workflows with (A) five and (B) ten operations.}}
    \label{exp:numpy_q}
\end{figure}

\begin{table}
    \centering
    \footnotesize
    \begin{tabularx}{\columnwidth}{l|ll}
    \hline
         & \textbf{Image Pipeline} & \textbf{Relational Pipeline} \\
    \hline
        1 & Resize to (416,416) & Inner Join on 'tconst' \\
        2 & Increase Luminosity & Filter Columns with NaN Value\\
        3  & Rotate $90^o$ &  Add Two Columns \\
        4 & Horizontal Flip  & One-Hot Encode on "genres" \\
        5 & Evaluate Lime on YOLOv4 & Add Constant to One Column \\
    \end{tabularx}
    \caption{Summary of Image and Relational Workflows}
    \label{table:pipelines}
\end{table}

We orchestrated \rev{five} experiments on different workflows to evaluate query latency - two on hand-generated workflows, \rev{one on a ResNet block \cite{resnet}} and two on randomly generated \texttt{numpy} workflows. The wall-clock latency time to the query result is measured, i.e., the time delay between the users issuing a query to \sys and receiving a response. Each \texttt{query\_cells} value is a randomly selected fixed-sized cell range. For cells within \texttt{query\_cells}, we query for the forward lineage through the full workflow pipeline. For brevity, backward query experiments are not shown in the paper - those results have similar trends as the forward query results since the algorithm applied are identical.

The first two experiments are over hand-generated workflows to match task-specific cases: one for computer vision model debugging and another for relational data pre-processing. Each of these workflows has five steps, including a mix of data-dependent and data-independent operations, as summarized in Table \ref{table:pipelines}. \rev{The Resnet workflow contains seven steps of a ResNet block during machine learning inference.} The image \rev{and ResNet} workflow is performed over a frame of the VIRAT dataset, and the relational workflow is performed over \rev{the full table of} the IMDB title.basics and title.episode tables. These are identical to data tables used in the compression storage experiments. 
 
For the two randomly-generated experiments, \texttt{numpy} operations are randomly selected from an initial list of 76 \texttt{numpy} operations\cite{extended_experiments}. This list includes selected operations from Section \ref{sec:np_coverage} that can intake and output a single \texttt{float\_64} arrays. Twenty workflows are generated using this list for each experiment with five and ten chained operations, respectively. The workflows operate on an initial array with randomly generated values and 100,000 cells. We use \texttt{tracked\_cells} to annotate the contribution lineage. 

For each experiment, the query latency time of our compression table and query algorithm in \sys is compared with the Parquet, Parquet-GZip, and Turbo-RC compressed tables and basic join queries, \rev{as well as vectorized \texttt{numpy} operations over the Array table. For the latter, we evaluated the equality condition (\texttt{==}) between the Array tables and the requested query (in array format). We batched the query with a batch size of 1000 to avoid out-of-memory issues.} For completeness, we included two additional baselines in the randomly generated workflows with five operations - the uncompressed table (Raw) and a version of our query algorithm where the final merge step is skipped (\sys-NoMerge). \rev{The maximize time to complete each query was set to 3 hours (10800 seconds).}

Figure \ref{exp:custom_q} shows the result of latency query experiments over the two hand-generated \rev{and ResNet} workflows. We define query selectivity by the percentage of cells from the initial dataset. The main observation is that \sys performed better than baselines over all queries \rev{except the most selective ones over the image workflow}, achieving up to \rev{1500x} less latency than the following best baseline. The main reason for this improvement is that fewer rows are queried due to the optimizations on our compression and query algorithms.\rev{The Array baseline performed the worst of all baselines and did not complete for less selective queries and queries over the relational workflow. This suggests the vectorized operation does not scale well with data and query size.} Turbo-RC has significant overhead because of decompression, making it highly unsuitable for more selective queries. Parquet and Parquet-GZip performed similarly; latency time increased  with the selectivity. The most significant benefit of \sys is on queries with low selectivity, with latency time increasing slower as the number of initial cells increased. Lastly, \sys had the \rev{lowest latency on the ResNet workflow, primarily because the structure of operations in the machine learning inference operations are extremely regular, and \FMain could compress such structures very efficiently.} 

Figure \ref{exp:numpy_q} shows the result of latency query experiments over the \texttt{numpy} workflows. Minimum and maximum latency times are shown with an interval bar. In general, the trends of the previous experiments hold, with \sys performing close to or better than baselines. \rev{Array's relative performance improved on more selective queries. We suspect that the smaller overhead of ingesting the array accounted for this difference on those queries.} The latency time of \sys was relatively higher, with only up to \rev{20x} improvement over the following best baseline. Again, this is accounted for by how much \FMain can compress each operation's lineage. All compression algorithms demonstrated a considerable variation in latency. The minimum/maximum query latency times were two orders of magnitude smaller/larger than the average time, showing that the specific operations captured significantly impacted query latency. 

The Raw queries were worse on very selective queries but were on par with Parquet and Parquet-GZip on less selective ones. \sys-NoMerge was worse than \sys over all queries, showing that the merge step significantly improves query performance with minimal overhead. There is an average of about a 4x increase in latency time on the pipelines with ten operations compared to the ones with five operations. The non-proportional increase is due to the complexity of the pipeline at ten operations. Despite this increase, the average query time is still significantly shorter using \sys.

\subsection{Reuse and Compression Over \texttt{numpy} API} 

\begin{table}
    \centering
    \footnotesize
    \begin{tabularx}{1\columnwidth}{lllllllll}
    \hline
        \textbf{Op.} & \textbf{Tot.} & \multicolumn{2}{c}{\textbf{\FMain}} & \multicolumn{2}{c}{\textbf{\texttt{dim\_sig}}}  &  \multicolumn{2}{c}{\textbf{\texttt{gen\_sig}}}   & \textbf{Error}   \\
         &  & Abs & \% & Abs & \% & Abs & \% & Abs  \\
    \hline
        element & 75 & 75 & 100 & 75 & 100 & 75 & 100 & 0  \\
        complex & 61 & 55 & 90.1 & 51 & 83.6 & 24 & 39.3& 1 \\
        total  & 136 & 130 & 95.5 & 126 & 92.6 & 99 & 72.7  & 1 \\
    \end{tabularx}
    \caption{\texttt{numpy} API Operations Covered by Compression and Reuse}
    \label{tab:np_coverage}
\end{table}

\label{sec:np_coverage} We quantitatively evaluate the coverage of \sys's optimization over 136 different operations in \texttt{numpy}. These optimizations include \FMain and automatic reuse prediction.

The operations consist of all functions in \texttt{numpy}'s API that meets the criteria of currently supported operation signatures in \texttt{tracked\_cells}: they (1) can intake and output \texttt{float\_64} arrays, and (2) can intake scalar-only arguments outside of \texttt{float\_64} arrays. They represent a diverse range of typical array manipulations and are a super-set of the operations used in the \texttt{numpy} query experiments in Section \ref{sec:exp-query} \cite{extended_experiments}.

\sys evaluated compression and reuse over 20 runs of each operation. The generated lineage is automatically compressed with \FMain, and the automatic reuse prediction algorithm is applied. For compression, we tallied the number of operations that were compressed to $<0.5$ of the original raw CSV file. For reuse, we evaluate the number of operations where \sys successfully discovers and stores shape-based and generalized mappings.

We divided the \texttt{numpy} operations into two categories - element-wise operation and other more complex patterns. Table \ref{tab:np_coverage} summarizes the results of our evaluation for both categories. This table shows that the coverage of \FMain compression is significant, and 95.5\% of all operations are compressed. This is because \texttt{numpy} operations are generally structured and value-independent. Our compression covered any operation covered by reuse and four additional operations. These results show that traditional array manipulation workflows are highly suited to lineage compression in \sys.

Automatic reuse prediction had less coverage than compression but performed exceptionally well over the \texttt{numpy} library. As the table shows, over half of our results are over element-wise operations. These operations were well-suited for reuse; \sys was able to generate shape-based and generalized mapping over all of those lineage tables. Mapping coverage over complex operations was slightly lower but still significant; shape and generalized reuse were generated for 86.8\% and 39.3\% of operations, respectively. This represents a significant reduction of capture latency compared to not using a lineage index such as \sys since retrieving each mapping has negligible latency. In contrast, naive lineage capture can be extremely costly.

One reuse misprediction error occurred, where \sys generated a wrong generalized mapping for the operation \texttt{cross}. This is because \texttt{cross} has different lineage patterns depending on the size of the second dimension. Mispredictions are the downside of setting $m$ as a low value in our prediction algorithm. 



\subsection{Compression over Data Science Workflows}

\begin{table}
    \centering
    \footnotesize
    \begin{tabularx}{1\columnwidth}{lllll}
    \hline
        \textbf{Dataset} & \textbf{Total Op.} & \multicolumn{2}{c}{\textbf{Compressible Op.}}  &  \textbf{Longest Chain}  \\
         &   & Abs & (\%) & \\
    \hline\
        Flight & 54.9 $\pm$ 38.8 & 40.5 $\pm$ 27.6 & 76.3 $\pm$ 11.0 & 16.4 $\pm$ 13.3 \\
        Netflix & 58.3 $\pm$ 36.3 & 40 $\pm$ 27.2 & 66.9 $\pm$ 9.2 & 14.2 $\pm$ 9.0 \\
        Total  & 56.6 $\pm$ 36.6 & 40.25 $\pm$ 26.7 & 71.6 $\pm$ 11.0 & 15.3 $\pm$ 11.12 \\
    \end{tabularx}
    \caption{Qualitative Estimate of Compressible Operations and Longest Operation Chain in Kaggle Workflows}
    \label{tab:coverage}
\end{table}

A very typical application of array operations is in data science workflows. We evaluated the coverage of \FMain over these workflows to ground the applicability of a lineage index in real-world scenarios. However, characterizing these workflows is difficult \cite{vamsa}, and to the best of our knowledge, there is currently no method to capture lineage over these workflows completely.

Instead of quantitative evaluation, 20 data science Jupyter notebooks were manually evaluated for coverage \cite{extended_experiments}. These notebooks were selected from the ``Trending'' code page for two popular Kaggle datasets: 2015 Flight Delays and Cancellations \cite{kaggleflight} and Netflix Movies and TV Shows \cite{kagglenetflix}. The Flight dataset contains 31 attributes and 5819079 rows and records flight data within the USA in 2015. The Netflix dataset contains 12 attributes and 8807 rows and records all content available on Netflix as of mid-2021. Notebooks were selected to span the spectrum between data exploration and machine learning predictions.

Coverage is evaluated as follows. Each operation with an input and output array is checked; visualization and plotting operations were excluded for simplicity. We considered an operation compressible if its estimated lineage matches one of the three patterns defined in Section \ref{sec:compression}, i.e., if its lineage contains continuous rectangular ranges in input cells, absolute output cells, or output cells after a relative transformation. Operations that did not explicitly match these patterns may still be compressible but were not counted. Operations appearing in for-loops over array indices were only considered once since they all were observed to have simple equivalence to similar operations that do not need such control structures.

The summary results of this experiment are shown in Table \ref{tab:coverage}. Qualitatively, we found that workflows that focused more on data exploration had higher operation counts, lower compressible operations, and shorter chains than machine learning workflows. We estimate that the distribution of data exploration and machine learning workflows accounted for the difference in results between the two datasets rather than the specific dataset properties.

The most significant result is the high coverage of compressible operations over these workflows(76.3$\pm$ 11.0\% on the Flight dataset and 66.9$\pm$ 9.2\% on the Netflix dataset). This suggests that the lineage of these operations would be significantly compressed with \FMain-GZip, achieving significant storage and query cost reduction when ingested into \sys. Most incompressible operations were value-filter operations, which do not have simple exploitable lineage patterns. 

During the inspection, each workflow's longest chained operation length was counted. The summary results of this count are in Table \ref{tab:coverage}. Across both datasets, we found that this length has less variation between workflows than the total number of operations and is not directly correlated with workflow length. The average length is 15.3$\pm$ 11.12, giving us an estimate of the upper bound of joins for lineage queries in tabular data workflows.

\section{Related Work}
To our knowledge, automatic compression has yet to be integrated into existing lineage systems for data science. For example, the Ground~\cite{hellerstein2017ground} system uses a Postgres SQL engine with a row-oriented data organization. Similarly, Smoke uses an in-memory data structure to store and index lineage for visualization applications~\cite{psallidas2018smoke}.


Prior work has been done on provenance compression of non-array structures, mainly in graph representation \cite{deutch2020hypothetical, heinis_2008}. Multiple works explore identifying and combining common nodes in a provenance graph \cite{xie2011compressing, xie2012hybrid, chapman2008efficient}. Anand et al. introduces range and subsequence reduction \cite{nested}. Compression has also been studied in network lineages ~\cite{chen2017distributed, zhou2012distributed}. Finally, algorithms have been presented for lossy compression of provenance through sampling \cite{why-not-2020} and minimizing graph distance \cite{prox}. Previous approaches do not exploit the numerical regularity of array indices in typical array-programming applications. We also leverage existing ideas from columnar compression (our baseline)~\cite{abadi2006integrating} and in-situ query processing~\cite{potti2015daq} to build our query processing framework.

Since our focus is on array programs, there is also a tangentially related field of machine learning workflow management.
Initial prototypes such as HELIX~\cite{xin2018helix}, Alpine Meadow~\cite{shang2109alpine}, MLFlow~\cite{zaharia2018accelerating}, and the Collaborative Optimizer~\cite{derakhshan2020optimizing}  rely on coarse-grained
lineage tracing at the level of entire machine learning programs.
Such systems are helpful in versioning and dataset discovery. However, they do not track contribution lineage and fail to give fine-grained information about how individual data values, i.e., individual array elements, contribute to final results. 

\bibliographystyle{abbrv}
\bibliography{refs}

\begin{thebibliography}{10}

\bibitem{pqtbest}
Athena compression support.

\bibitem{extended_experiments}
Detailed coverage experiment results.
\newblock {https://github.com/j2zhao/DSLog-Coverage}, journal={Github}, year={2023}, month={May}.

\bibitem{range_join}
Range join optimization.
\newblock https://docs.databricks.com/optimizations/range-join.html.

\bibitem{IMDb_noncommercial_datasets}
Imdb non-commercial datasets.
\newblock https://developer.imdb.com/non-commercial-datasets/, 2022.

\bibitem{abadi2006integrating}
D.~Abadi, S.~Madden, and M.~Ferreira.
\newblock Integrating compression and execution in column-oriented database systems.
\newblock In {\em Proceedings of the 2006 ACM SIGMOD international conference on Management of data}, pages 671--682, 2006.

\bibitem{prox}
E.~Ainy, P.~Bourhis, S.~B. Davidson, D.~Deutch, and T.~Milo.
\newblock Approximated summarization of data provenance.
\newblock In {\em Proceedings of the 24th ACM International on Conference on Information and Knowledge Management}, CIKM '15, page 483–492, New York, NY, USA, 2015. Association for Computing Machinery.

\bibitem{nested}
M.~Anand, S.~Bowers, T.~McPhillips, and B.~Lud{\"a}scher.
\newblock Efficient provenance storage over nested data collections.
\newblock In {\em Proceedings of the 12th International Conference on Extending Database Technology}, Proceedings of the 12th International Conference on Extending Database Technology: Advances in Database Technology, EDBT'09, pages 958--969, 2009.
\newblock Copyright: Copyright 2009 Elsevier B.V., All rights reserved.; 12th International Conference on Extending Database Technology: Advances in Database Technology, EDBT'09 ; Conference date: 24-03-2009 Through 26-03-2009.

\bibitem{kagglenetflix}
S.~Bansal.
\newblock Netflix movies and tv shows.
\newblock {https://www.kaggle.com/datasets/shivamb/netflix-shows}, Sep 2021.

\bibitem{boehm2016systemml}
M.~Boehm, M.~W. Dusenberry, D.~Eriksson, A.~V. Evfimievski, F.~M. Manshadi, N.~Pansare, B.~Reinwald, F.~R. Reiss, P.~Sen, A.~C. Surve, et~al.
\newblock Systemml: Declarative machine learning on spark.
\newblock {\em Proceedings of the VLDB Endowment}, 9(13):1425--1436, 2016.

\bibitem{chapman2008efficient}
A.~P. Chapman, H.~V. Jagadish, and P.~Ramanan.
\newblock Efficient provenance storage.
\newblock In {\em Proceedings of the 2008 ACM SIGMOD international conference on Management of data}, pages 993--1006, 2008.

\bibitem{chen2017distributed}
C.~Chen, H.~T. Lehri, L.~Kuan~Loh, A.~Alur, L.~Jia, B.~T. Loo, and W.~Zhou.
\newblock Distributed provenance compression.
\newblock In {\em Proceedings of the 2017 ACM International Conference on Management of Data}, pages 203--218, 2017.

\bibitem{cheney2009provenance}
J.~Cheney, L.~Chiticariu, and W.-C. Tan.
\newblock {\em Provenance in databases: Why, how, and where}.
\newblock Now Publishers Inc, 2009.

\bibitem{ciucanu2015worst}
R.~Ciucanu and D.~Olteanu.
\newblock Worst-case optimal join at a time.
\newblock Technical report, Technical report, Oxford, 2015.

\bibitem{cloud2017configurable}
C.~Cloud.
\newblock A configurable experimental environment for large-scale cloud research. chameleoncloud. org. retrieved, 2022.

\bibitem{derakhshan2020optimizing}
B.~Derakhshan, A.~Rezaei~Mahdiraji, Z.~Abedjan, T.~Rabl, and V.~Markl.
\newblock Optimizing machine learning workloads in collaborative environments.
\newblock In {\em Proceedings of the 2020 ACM SIGMOD International Conference on Management of Data}, pages 1701--1716, 2020.

\bibitem{deutch2020hypothetical}
D.~Deutch, Y.~Moskovitch, and N.~Rinetzky.
\newblock Hypothetical reasoning via provenance abstraction, 2020.

\bibitem{hanzo2002turbo}
L.~Hanzo, T.~H. Liew, and B.~L. Yeap.
\newblock {\em Turbo coding, turbo equalisation and space-time coding}.
\newblock John Wiley \& Sons, 2002.

\bibitem{resnet}
K.~He, X.~Zhang, S.~Ren, and J.~Sun.
\newblock Deep residual learning for image recognition.
\newblock {\em CoRR}, abs/1512.03385, 2015.

\bibitem{heinis_2008}
T.~Heinis and G.~Alonso.
\newblock Efficient lineage tracking for scientific workflows.
\newblock In {\em Proceedings of the 2008 ACM SIGMOD International Conference on Management of Data}, SIGMOD '08, page 1007–1018, New York, NY, USA, 2008. Association for Computing Machinery.

\bibitem{hellerstein2017ground}
J.~M. Hellerstein, V.~Sreekanti, J.~E. Gonzalez, J.~Dalton, A.~Dey, S.~Nag, K.~Ramachandran, S.~Arora, A.~Bhattacharyya, S.~Das, et~al.
\newblock Ground: A data context service.
\newblock In {\em CIDR}. Citeseer, 2017.

\bibitem{interlandi2015titian}
M.~Interlandi, K.~Shah, S.~D. Tetali, M.~A. Gulzar, S.~Yoo, M.~Kim, T.~Millstein, and T.~Condie.
\newblock Titian: Data provenance support in spark.
\newblock In {\em Proceedings of the VLDB Endowment International Conference on Very Large Data Bases}, volume~9, page 216. NIH Public Access, 2015.

\bibitem{why-not-2020}
S.~Lee, B.~Lud{\"{a}}scher, and B.~Glavic.
\newblock Approximate summaries for why and why-not provenance (extended version).
\newblock {\em CoRR}, abs/2002.00084, 2020.

\bibitem{vamsa}
M.~H. Namaki, A.~Floratou, F.~Psallidas, S.~Krishnan, A.~Agrawal, and Y.~Wu.
\newblock Vamsa: Tracking provenance in data science scripts.
\newblock {\em CoRR}, abs/2001.01861, 2020.

\bibitem{nikolic2014linview}
M.~Nikolic, M.~Elseidy, and C.~Koch.
\newblock Linview: incremental view maintenance for complex analytical queries.
\newblock In {\em Proceedings of the 2014 ACM SIGMOD international conference on Management of data}, pages 253--264, 2014.

\bibitem{kaggleflight}
D.~of~Transportation.
\newblock 2015 flight delays and cancellations.
\newblock {https://www.kaggle.com/datasets/usdot/flight-delays}, Feb 2017.

\bibitem{oh2011large}
S.~Oh, A.~Hoogs, A.~Perera, N.~Cuntoor, C.-C. Chen, J.~T. Lee, S.~Mukherjee, J.~Aggarwal, H.~Lee, L.~Davis, et~al.
\newblock A large-scale benchmark dataset for event recognition in surveillance video.
\newblock In {\em CVPR 2011}, pages 3153--3160. IEEE, 2011.

\bibitem{olteanu2011factorisation}
D.~Olteanu.
\newblock On factorisation of provenance polynomials.
\newblock In {\em In TaPP}. Citeseer, 2011.

\bibitem{Petsiuk2021drise}
V.~Petsiuk, R.~Jain, V.~Manjunatha, V.~I. Morariu, A.~Mehra, V.~Ordonez, and K.~Saenko.
\newblock Black-box explanation of object detectors via saliency maps.
\newblock In {\em Computer Vision and Pattern Recognition (CVPR)}, 2021.

\bibitem{phani2021lima}
A.~Phani, B.~Rath, and M.~Boehm.
\newblock Lima: Fine-grained lineage tracing and reuse in machine learning systems.
\newblock In {\em Proceedings of the 2021 International Conference on Management of Data}, pages 1426--1439, 2021.

\bibitem{pibiri2020techniques}
G.~E. Pibiri and R.~Venturini.
\newblock Techniques for inverted index compression.
\newblock {\em ACM Computing Surveys (CSUR)}, 53(6):1--36, 2020.

\bibitem{potti2015daq}
N.~Potti and J.~M. Patel.
\newblock Daq: a new paradigm for approximate query processing.
\newblock {\em Proceedings of the VLDB Endowment}, 8(9):898--909, 2015.

\bibitem{psallidas2018smoke}
F.~Psallidas and E.~Wu.
\newblock Smoke: Fine-grained lineage at interactive speed.
\newblock {\em arXiv preprint arXiv:1801.07237}, 2018.

\bibitem{raasveldt2019duckdb}
M.~Raasveldt and H.~M{\"u}hleisen.
\newblock Duckdb: an embeddable analytical database.
\newblock In {\em Proceedings of the 2019 International Conference on Management of Data}, pages 1981--1984, 2019.

\bibitem{rezig2020dagger}
E.~K. Rezig, L.~Cao, G.~Simonini, M.~Schoemans, S.~Madden, N.~Tang, M.~Ouzzani, and M.~Stonebraker.
\newblock Dagger: a data (not code) debugger.
\newblock In {\em CIDR 2020, 10th Conference on Innovative Data Systems Research, Amsterdam, The Netherlands, January 12-15, 2020, Online Proceedings}, 2020.

\bibitem{ribeiro2016should}
M.~T. Ribeiro, S.~Singh, and C.~Guestrin.
\newblock " why should i trust you?" explaining the predictions of any classifier.
\newblock In {\em Proceedings of the 22nd ACM SIGKDD international conference on knowledge discovery and data mining}, pages 1135--1144, 2016.

\bibitem{sellam2019deepbase}
T.~Sellam, K.~Lin, I.~Huang, M.~Yang, C.~Vondrick, and E.~Wu.
\newblock Deepbase: Deep inspection of neural networks.
\newblock In {\em Proceedings of the 2019 International Conference on Management of Data}, pages 1117--1134, 2019.

\bibitem{shang2019democratizing}
Z.~Shang, E.~Zgraggen, B.~Buratti, F.~Kossmann, P.~Eichmann, Y.~Chung, C.~Binnig, E.~Upfal, and T.~Kraska.
\newblock Democratizing data science through interactive curation of ml pipelines.
\newblock In {\em Proceedings of the 2019 international conference on management of data}, pages 1171--1188, 2019.

\bibitem{shang2109alpine}
Z.~Shang, E.~Zgraggen, and T.~Kraska.
\newblock Alpine meadow: A system for interactive automl.

\bibitem{suh2004secure}
G.~E. Suh, J.~W. Lee, D.~Zhang, and S.~Devadas.
\newblock Secure program execution via dynamic information flow tracking.
\newblock {\em ACM Sigplan Notices}, 39(11):85--96, 2004.

\bibitem{tiwari2009complete}
M.~Tiwari, H.~M. Wassel, B.~Mazloom, S.~Mysore, F.~T. Chong, and T.~Sherwood.
\newblock Complete information flow tracking from the gates up.
\newblock In {\em Proceedings of the 14th international conference on Architectural support for programming languages and operating systems}, pages 109--120, 2009.

\bibitem{vartak2018mistique}
M.~Vartak, J.~M. F.~da Trindade, S.~Madden, and M.~Zaharia.
\newblock Mistique: A system to store and query model intermediates for model diagnosis.
\newblock In {\em Proceedings of the 2018 International Conference on Management of Data}, pages 1285--1300, 2018.

\bibitem{widom2004trio}
J.~Widom.
\newblock Trio: A system for integrated management of data, accuracy, and lineage.
\newblock Technical report, Stanford InfoLab, 2004.

\bibitem{xie2012hybrid}
Y.~Xie, D.~Feng, Z.~Tan, L.~Chen, K.-K. Muniswamy-Reddy, Y.~Li, and D.~D. Long.
\newblock A hybrid approach for efficient provenance storage.
\newblock In {\em Proceedings of the 21st ACM international conference on Information and knowledge management}, pages 1752--1756, 2012.

\bibitem{xie2011compressing}
Y.~Xie, K.-K. Muniswamy-Reddy, D.~D. Long, A.~Amer, D.~Feng, and Z.~Tan.
\newblock Compressing provenance graphs.
\newblock In {\em 3rd USENIX Workshop on the Theory and Practice of Provenance (TaPP 11)}, 2011.

\bibitem{xin2018helix}
D.~Xin, S.~Macke, L.~Ma, J.~Liu, S.~Song, and A.~Parameswaran.
\newblock Helix: Holistic optimization for accelerating iterative machine learning.
\newblock {\em arXiv preprint arXiv:1812.05762}, 2018.

\bibitem{zaharia2018accelerating}
M.~Zaharia, A.~Chen, A.~Davidson, A.~Ghodsi, S.~A. Hong, A.~Konwinski, S.~Murching, T.~Nykodym, P.~Ogilvie, M.~Parkhe, et~al.
\newblock Accelerating the machine learning lifecycle with mlflow.
\newblock {\em IEEE Data Eng. Bull.}, 41(4):39--45, 2018.

\bibitem{zhang2017diagnosing}
Z.~Zhang, E.~R. Sparks, and M.~J. Franklin.
\newblock Diagnosing machine learning pipelines with fine-grained lineage.
\newblock In {\em Proceedings of the 26th International Symposium on High-Performance Parallel and Distributed Computing}, pages 143--153, 2017.

\bibitem{zhou2012distributed}
W.~Zhou, S.~Mapara, Y.~Ren, Y.~Li, A.~Haeberlen, Z.~Ives, B.~T. Loo, and M.~Sherr.
\newblock Distributed time-aware provenance.
\newblock {\em Proceedings of the VLDB Endowment}, 6(2):49--60, 2012.

\bibitem{zhu2011tainteraser}
D.~Zhu, J.~Jung, D.~Song, T.~Kohno, and D.~Wetherall.
\newblock Tainteraser: Protecting sensitive data leaks using application-level taint tracking.
\newblock {\em ACM SIGOPS Operating Systems Review}, 45(1):142--154, 2011.

\end{thebibliography}


\end{document}